\documentclass[%
 reprint,
 twocolumn,
 superscriptaddress,
 amsmath,amssymb,
]{revtex4-2}

\usepackage{graphicx}
\usepackage{dcolumn}
\usepackage{bm}
\usepackage{soul}
\usepackage{tabularx} 
\usepackage[dvipsnames]{xcolor}
\usepackage[english]{babel}
\usepackage{hyperref}
\usepackage{rotating}
\usepackage{float}
\usepackage{multirow}
\usepackage{rotating}

\newcommand{\commentout}[1]{\ignorespaces}

\begin{document}

\title{Deep-learning-powered data analysis in plankton ecology}

\author{Harshith Bachimanchi}
\thanks{Equal authorship. Names are listed in alphabetical order by the last names.}
\affiliation{Department of Physics, University of Gothenburg, Sweden}

\author{Matthew I. M. Pinder}
\thanks{Equal authorship. Names are listed in alphabetical order by the last names.}
\affiliation{Department of Marine Sciences, University of Gothenburg, Sweden}

\author{Chloé Robert}
\thanks{Equal authorship. Names are listed in alphabetical order by the last names.}
\affiliation{Department of Marine Sciences, University of Gothenburg, Sweden}

\author{Pierre De Wit}
\affiliation{Department of Marine Sciences, University of Gothenburg, Sweden}

\author{Jonathan Havenhand}
\affiliation{Department of Marine Sciences, University of Gothenburg, Sweden}

\author{Alexandra Kinnby}
\affiliation{Department of Marine Sciences, University of Gothenburg, Sweden}

\author{Daniel Midtvedt}
\affiliation{Department of Physics, University of Gothenburg, Sweden}

\author{Erik Selander}
\affiliation{Department of Biology, Lund University, Sweden}

\author{Giovanni Volpe}
\affiliation{Department of Physics, University of Gothenburg, Sweden}

\date{\today}

\begin{abstract}
The implementation of deep learning algorithms has brought new perspectives to plankton ecology. 
Emerging as an alternative approach to established methods, deep learning offers objective schemes to investigate plankton organisms in diverse environments. 
We provide an overview of deep-learning-based methods including detection and classification of phyto- and zooplankton images, foraging and swimming behaviour analysis, and finally ecological modelling. 
Deep learning has the potential to speed up the analysis and reduce the human experimental bias, thus enabling data acquisition at relevant temporal and spatial scales with improved reproducibility. 
We also discuss shortcomings and show how deep learning architectures have evolved to mitigate imprecise readouts. 
Finally, we suggest opportunities where deep learning is particularly likely to catalyze plankton research. 
The examples are accompanied by detailed tutorials and code samples that allow readers to apply the methods described in this review to their own data \cite{bachimanchi_planktonreview_2023}.
\end{abstract}
\maketitle

\section{Introduction}

The monitoring of marine organisms' composition and abundance plays a pivotal role in evaluating environmental conditions and gaining insights into ecological processes and interactions \cite{Racault2014}.
One group of particular interest is marine plankton, a collection of organisms defined by their inability to swim against the oceanic currents \cite{Hensen1887}.
Plankton encompasses diverse organisms, from unicellular bacteria and protozoans to eumetazoans such as small crustaceans and gelatinous life forms.
Given their diverse size (ranging from micrometre to metre scales), rapid turnover, and displacement with ocean currents, these organisms are challenging to study at appropriate temporal and spatial scales. 
While studies of microplankton have historically been performed using microscopy and chemical methods, such as pigment analysis and isotope labelling \cite{Butterwick1982, kiorboe2009mechanistic}, advances in such methods, alongside new developments in molecular biology such as genetic barcoding, have increased the taxonomic resolution in plankton monitoring and revealed a previously hidden diversity in many groups.
However, most methods are still too time-consuming and labour-intensive to resolve the plankton community at relevant spatial and temporal scales.
Recent advances in machine learning, such as deep-learning-based methods, hold promise to improve this situation. In particular, automatic classification of organisms from {\it in situ} imaging systems allows high resolution in both time and space~\cite{orenstein2020scripps}.
Moreover, developments in deep learning microscopy provide single cell resolution in food web interactions, such as microzooplankton grazing and behaviour, which have previously been hard to obtain \cite{bachimanchi2022microplankton}.  
In this current state review, we aim to delve into the application of deep learning methods in plankton studies, with a particular focus on utilising image data. 
Our objective is to shed light on the potential of these methods as tools for researchers in the field of plankton ecology, especially for those who may not be familiar with deep learning techniques.

Plankton play a pivotal role in the Earth's ecosystem and especially in its carbon cycle.
Phytoplankton, for example, produce nearly half of the oxygen on Earth, performing around 40\% of the world's annual carbon fixation (30-50 petagrams)~\cite{Falkowski1994}.
Together with zooplankton as primary consumers, these form the base of most marine food webs.
Certain species of plankton can also pose hazards to human health, such as harmful algal bloom-forming taxa (e.g., toxin-producing diatoms and dinoflagellates~\cite{Bates2018}).
In ecological terms, the composition and abundance of certain species is used to determine ecological status~\cite{Racault2014, HELCOM}, or to obtain insights into past conditions (e.g., using shifts in diatom composition from sediment cores to make inferences about the climate in the past few millennia~\cite{mackay2003approaches}).

\begin{figure*}[pt]
    \label{glossary}
    \fbox{\begin{minipage}{17cm}
    \begin{flushleft}
    \textbf{Box 1. Glossary of deep learning terms as used in this review}\\
    \textbf{Back-propagation algorithm}: A training algorithm for neural networks that adjusts weights based on error gradients. \\
    \textbf{Bounding box}: A rectangular box defined by latitude and longitude coordinates that enclose the object(s) of interest in an image. \\
    \textbf{Class}: A category or label assigned to data points. \\
    \textbf{CycleGAN}: A type of generative adversarial network designed to create realistic-looking images without paired input-output examples. \\
    \textbf{Data augmentation}: A technique to enhance the size and variety of a training dataset by applying transformations to the existing data to improve model generalization. \\
    \textbf{Deep learning}: A subfield of machine learning that models data using multiple layers of interconnected nodes. It is inspired by the structure and function of the biological brain. \\
    \textbf{Feed-forward network}: A neural network with connections that do not form cycles.\\
    \textbf{F1 score}: A metric combining precision and recall, defined as $2 * \frac{{\rm Precision} * {\rm Recall}}{{\rm Precision} + {\rm Recall}}$. \\
    \textbf{GNN (Graph Neural Network)}: Neural networks designed to operate on graphs, mathematical structures composed of nodes (vertices) and edges. \\
    \textbf{Human-in-the-loop}: An iterative process where a human provides feedback on a system's outputs, aiding in improving its performance. \\
    \textbf{Layers}: Sets of interconnected nodes or units in a neural network. Data is processed through each layer, undergoing transformations before moving to subsequent layers or producing an output. \\
    \textbf{Long short-term memory (LSTM)}: A type of recurrent neural network (RNN) architecture that is designed to recognize patterns over time intervals and has memory cells to store information for longer periods. \\
    \textbf{Multilayer perceptron (MLP)}: A type of feed-forward neural network with multiple layers of nodes. \\
    \textbf{Overfitting}: A modeling error that occurs when a machine learning model performs well on the training data but poorly on new, unseen data. \\
    \textbf{Precision}: The ratio of correctly predicted positive observations to the total predicted positives, defined as $\frac{{\rm TP}}{{\rm TP} + {\rm FP}}$. \\
    \textbf{R-CNN (Regional Convolutional Neural Network)}: A model used for object detection that scans an image in regions and uses CNNs to classify each region. \\
    \textbf{Recall}: The ratio of correctly predicted positive observations to all the observations in actual class, defined as $\frac{{\rm TP}}{{\rm TP} + {\rm FN}}$. \\
    \textbf{Reinforcement learning}: A learning paradigm where agents learn by interacting with an environment, receiving feedback through rewards or penalties based on their actions. \\
    \textbf{Specificity}: The ratio of correctly predicted negative observations to all the observations in actual negative class, defined as $\frac{{\rm TN}}{{\rm TN} + {\rm FP}}$. \\
    \textbf{Test data}: A subset of data used to evaluate the performance of a trained model, ensuring it works effectively on new, unseen examples. \\
    \textbf{Training dataset}: A set of data used to train and adjust the weights of a model. \\
    \textbf{Transfer learning}: A technique wherein a pre-trained model on a large dataset is adapted for a different but related task, often involving fine-tuning on a smaller, task-specific dataset. \\
    \textbf{Validation dataset}: A subset of data used to tune hyperparameters and prevent overfitting during the training process. \\
    \textbf{YOLO (You Only Look Once)}: A real-time object detection system that divides images into a grid and predicts bounding boxes and class probabilities simultaneously.
    \end{flushleft}
\end{minipage}}
\end{figure*}

To accomplish these goals, it is important to have reliable methods to identify, track, and quantify plankton species or groups. Unfortunately, this is complicated by the extraordinary diversity among plankton taxa.
For example, just one class of phytoplankton --- the diatoms --- contains an estimated number of species in the tens of thousands~\cite{malviya2016insights}, many of which are highly similar in appearance and require microscopic analysis by expert taxonomists to differentiate.
Molecular methods such as metabarcoding and environmental DNA (eDNA) drastically improves the taxonomic resolution and has revealed a previously overlooked diversity in plankton communities~\cite{bucklin2016metabarcoding,ruppert2019past}.

\begin{figure*}
    \centering
    \includegraphics[scale=0.85]{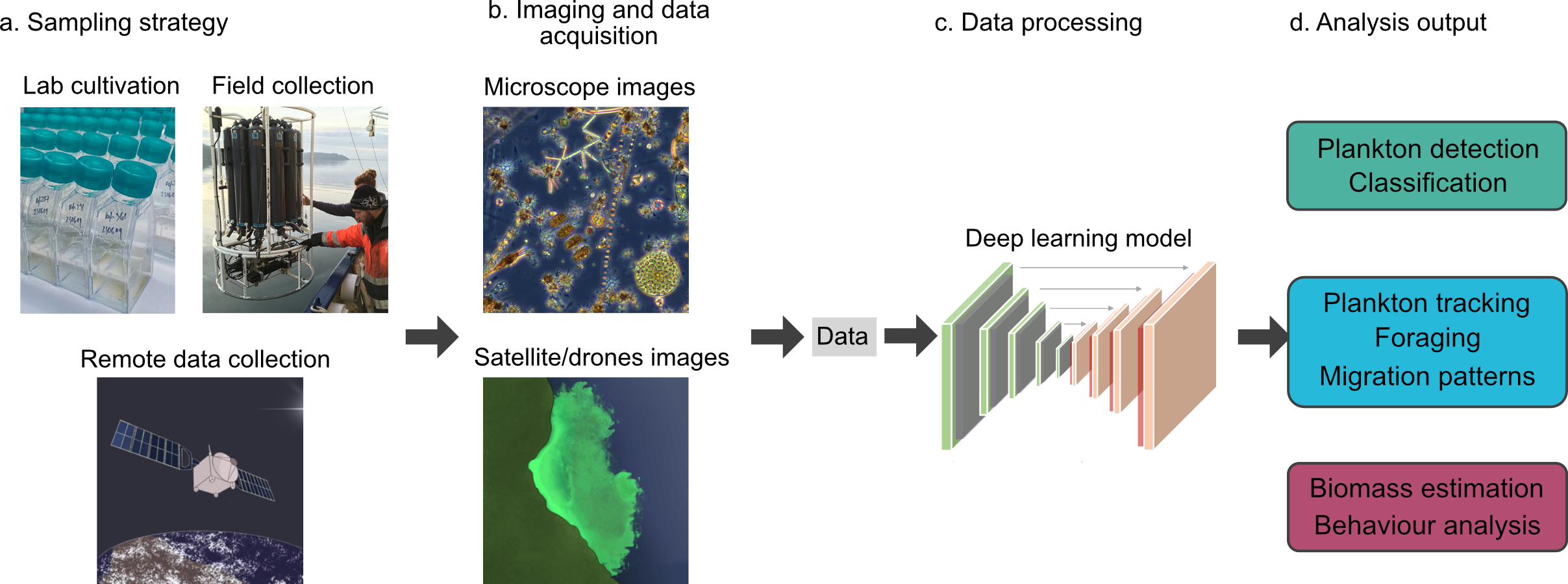}
    \caption{
    \label{figure1}
    Workflow of a deep learning analysis applied to plankton studies. 
    {\bf a} Different sampling strategies for \textit{in situ} and \textit{ex situ} data. Plankton can be cultivated in the laboratory or collected from the field. Remote-sensing data can also be collected via drones or satellites. 
    {\bf b} Images and data are obtained using microscopic analyses or satellite imaging. 
    {\bf c}  Data is processed using a deep learning network. 
    {\bf d}  The output of a deep learning analysis can be used for a wide range of applications.}
\end{figure*}

Deep learning has become increasingly popular in plankton studies.
This particular subset of machine learning harnesses neural networks to acquire knowledge and enhance performance in specific tasks through iterative learning from experience \cite{LeCun2015}.
Deep learning methods have been applied to many fields, such as natural language processing \cite{Young2018} and computer vision \cite{Chai2021}.
While an in-depth explanation of these networks is outside the scope of this paper, we will provide an overview of how these networks function, and recommend other reviews on the subject (such as those by LeCun \textit{et al.} \cite{LeCun2015} and Alzubaidi \textit{et al.} \cite{Alzubaidi2021}) to those readers interested in the finer details of deep learning.

As its name implies, a neural network's design is inspired by the brain, with an interconnected network of neurons.
The network to be trained is fed a set of input data ({\it training data}), most often labelled (e.g., a selection of plankton images identified by a taxonomist).
From these training data, the network learns to recognise certain features, and to associate these features or combinations thereof with given classifications.
The network's ability to recognise these features (and by extension, to classify the images) is then tested on a subset of the input dataset that the model has never seen before ({\it validation data}) to assess the model's performance.
Adjustments are made to the internal parameters of the network to minimise the error (i.e., the difference between the true results and the network predictions), and the process is repeated multiple times.
Finally, the model is tested on another unseen subset of the input dataset ({\it test data}) to provide the final statistics for the model's performance.
For tasks such as plankton classification and detection, the performance of the models can be evaluated using metrics such as: {\it precision}, i.e., the proportion of positive results identified that are truly positive; {\it recall} or {\it sensitivity}, i.e., the proportion of truly positive results that the model recognises as positive; {\it F1 score}, i.e., the harmonic mean of precision and recall; and {\it specificity}, i.e., the proportion of truly negative results that the model recognises as negative.

With this review, we aim to provide an overview of the key techniques used in deep learning, through the lens of their application in studies of plankton detection and classification (section II), plankton foraging and migration (section III), and plankton biomass estimation (section IV) (Fig.~\ref{figure1}).
We aim to demystify deep learning and inspire its use in future plankton projects.
As many field-specific terms are used throughout this review, we define important concepts in Box 1 (terms defined this way are italicised in the text at first use). 
We also show example schematics for some of the deep learning architectures (Figs.~\ref{figure2} and 3).
In addition, we provide Jupyter notebooks containing example code for three of the deep-learning-based tasks covered in this article: plankton segmentation and classification, plankton detection, and plankton trajectory linking.
The necessary code and plankton video data can be found in our GitHub repository~\cite{bachimanchi_planktonreview_2023}.
We encourage the readers to go through the tutorials and apply the techniques to their own plankton microscopy data.

\section{Plankton Detection and Classification}

Given the importance of plankton in the ecosystem, it is vital to be able to accurately detect and classify them.
Such tasks are non-trivial, as images of plankton may contain non-plankton particles, as well as multiple species of plankton that may have variable and sometimes largely overlapping morphologies.
Much of this work is traditionally performed manually, which constitutes a major bottleneck in analysis pipelines.
Therefore, emerging technologies such as metabarcoding~\cite{Abad2016} (the sequencing of taxon-specific DNA to identify which species are present) and deep learning methods present attractive alternatives, either individually or as complements to one another (as noted by Høye \textit{et al.} for the field of entomology~\cite{hoye2021deep}).
A major advantage of deep-learning-based approaches in this context is the potential for a more automated analysis after the initial setup.
While initial labelling of training images and training of the model may take some time, later analyses of plankton images do not require the repeated laboratory steps of, for example, a barcoding-based method.

The presence of numerous plankton species, numbering in the thousands, underscores a significant factor to consider when employing deep learning methods. One such consideration is the necessity for a substantial training dataset that encompasses a wide range of categories, often referred to as {\it classes} in the context of machine learning.
These would correspond to taxonomic classifications (e.g., species) in a plankton study.
Several curated plankton image datasets are available, which can be used for this purpose.
For example, the Woods Hole Oceanographic Institution Plankton dataset (WHOI-Plankton) contains over 3.4 million annotated plankton images belonging to 103 classes~\cite{orenstein2015whoi}.
In terms of more taxon-specific datasets, the Endless Forams dataset contains over 34,000 images of foraminifera classified to species level~\cite{hsiang_endless_2019}.

\begin{figure*}[pt]
    \centering
    \includegraphics[scale = 0.8, angle=0]{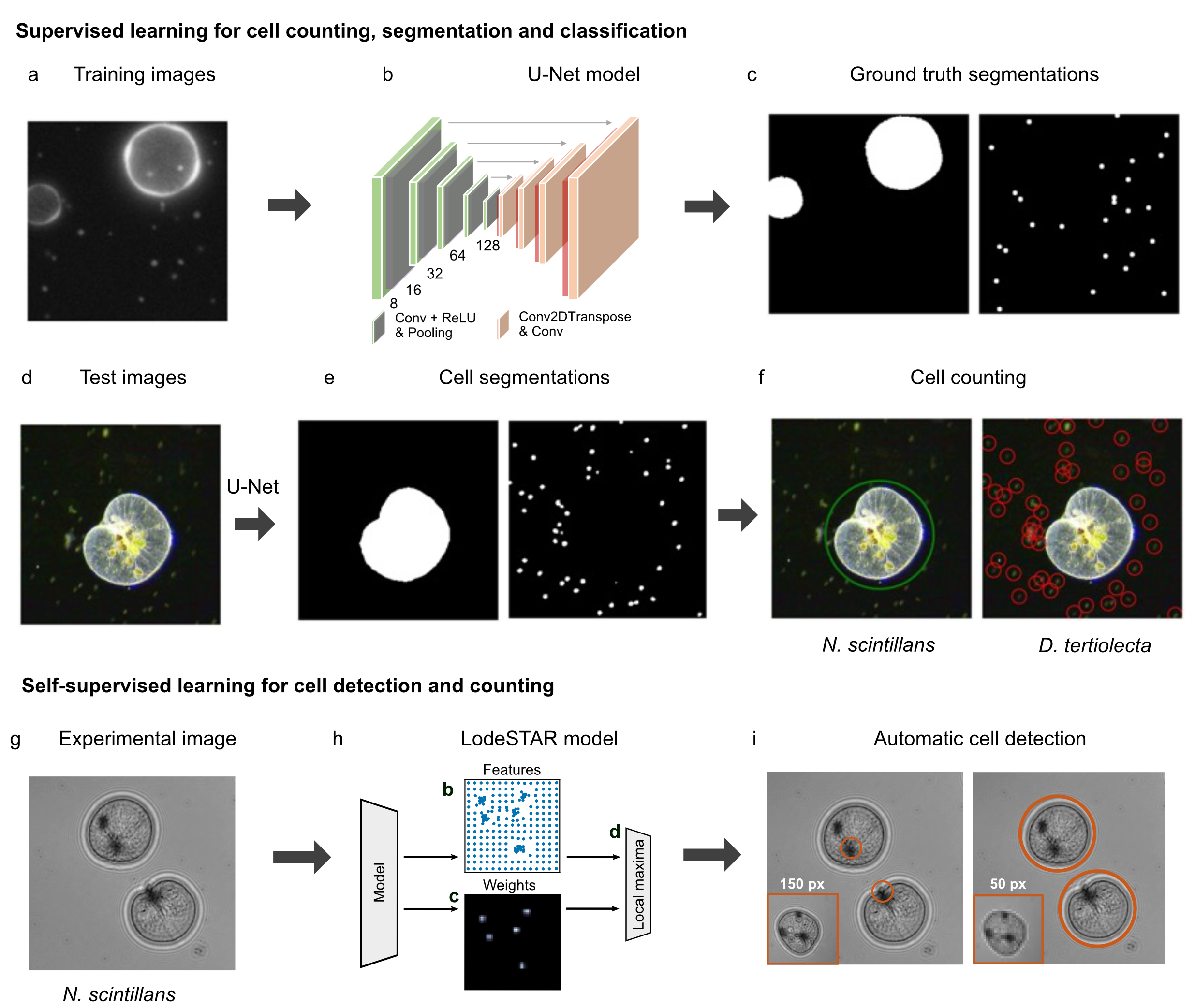}
    \caption{
    \label{figure2}
    {\bf Deep learning methods for cell counting, segmentation and classification.}
    {\bf a}-{\bf f} Example of a supervised method to detect, classify, and count plankton species {\it Noctiluca scintillans} and {\it Dunaliella tertiolecta}:
    {\bf a} Synthetic training images are generated by simulating plankton-like objects that carry similar spatial features to the original organisms.
    {\bf b} A U-Net model is trained to perform an image-to-image translation task to output {\bf c} two segmentation masks that correspond to the species {\it N. scinitillans} (left) and {\it D. tertiolecta} (right).
    {\bf d} The pre-trained U-Net model is tested on experimental images to generate {\bf e} the specific segmentation masks of {\it N. scintillans } (left) and {\it D. tertiolecta} (right).
    {\bf f} The segmentation masks are then analysed to predict the number of cells (cell counting) for each species in the experimental image.
    {\bf g}-{\bf i} Example of a self-supervised method to detect and count plankton.
    {\bf g} The experimental image of {\it N. scintillans} is analysed {\bf h} through a self-supervised object detection model, LodeSTAR~\cite{midtvedt_single-shot_2022}, to automatically detect the objects without any labelled training data.
    {\bf i} By controlling the spatial resolution of the input image, object centroids are controlled in the cell detection step.
    Figures {\bf g}-{\bf i} are adapted from Ref.\cite{midtvedt_single-shot_2022}, with some of the authors of this work being part of the cited reference.
    }
\end{figure*}

\begin{figure*}
    \centering
    \includegraphics[scale = 0.8, angle=0]{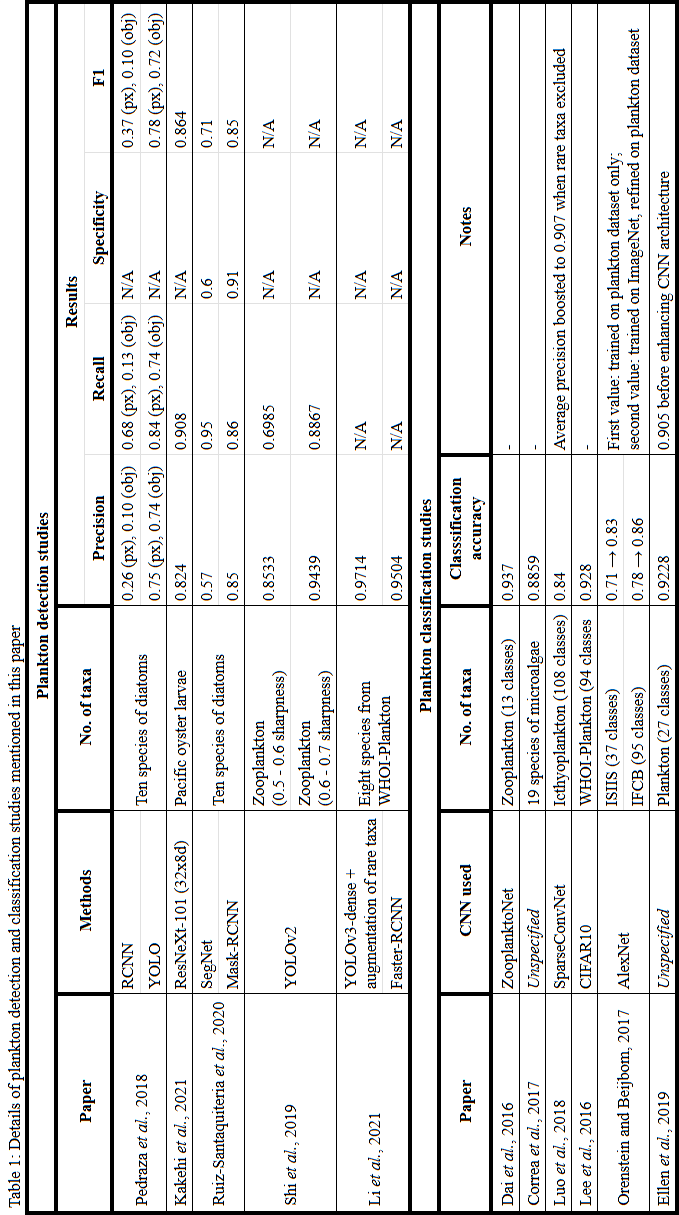}
    \caption{
    \label{table1}
    }
\end{figure*}

{\bf Plankton detection.} 
In order to classify plankton, they must first be detected in images or videos.
These can be obtained, for example, from a microscope in a lab context, or from imaging planktons moored in position or deployed on remotely operated vehicles such as the zooplankton-sensing autonomous glider Zooglider~\cite{Ohman2019}, or more integrated analysis and acquisition systems such as ZooScan~\cite{gorsky2010digital} and the Underwater Vision Profiler 6 (UVP6)~\cite{picheral2022underwater}.
Though most of these systems work favourably well, accurate plankton detection and classification can still be complicated by the issues such as motion blur in the acquired imagery data, or the presence of non-plankton objects such as aggregates and debris.
Therefore, sophisticated methods are required to detect true instances of plankton while rejecting non-plankton objects, which can be achieved through deep learning (Fig.~\ref{figure2}). (A summary of the studies in the following paragraphs are available in Table~\ref{table1}).

Two common approaches for object detection using deep learning are models of either the {\it R-CNN} (Regional Convolutional Neural Network) or {\it YOLO} (You Only Look Once) family. The R-CNN family of models (e.g., R-CNN, Fast R-CNN, Faster R-CNN, etc.) perform object detection and classification in a two-step process~\cite{Girshick2013}.
In the RCNN workflow, initially, regions of interest (ROIs) are generated, involving the cropping of individual plankton, followed by the merging of similar regions; subsequently, these resultant region proposals undergo classification.
The initial implementations of R-CNNs required processing of around 2,000 region proposals per image.
However, this was a substantial time bottleneck.
The more recent implementations are faster because this process has been significantly sped up; this improved performance means that Faster R-CNN can be used for real-time applications \cite{Ren2015}.
On the other hand, in YOLO models (e.g., YOLO, YOLOv2, YOLOv3, etc.), the detection and classification are performed across the whole image in a single step instead (hence the acronym), rather than first generating region proposals as in R-CNN, which further speeds up the process \cite{Redmon2016}.

Pedraza \textit{et al.}~\cite{Pedraza2018} applied both R-CNN and YOLO to microscopy images of ten diatom species in order to compare the two methods.
YOLO produced better results in this instance, with an average pixelwise recall of 0.84, versus an average of 0.68 for R-CNN.
In an intuitive sense, recall indicates the ability of a model to find all relevant detections.
The issues noted by Pedraza \textit{et al.}~\cite{Pedraza2018} regarding R-CNN were, objects being missed (i.e., false negatives), and {\it bounding boxes} being much larger than objects (i.e., false positives).
The latter also relates to the problem of multiple objects that are close together or overlapping being counted as a single object, which is problematic for obtaining accurate numbers.

Issues similar to those noted by Pedraza \textit{et al.}~\cite{Pedraza2018} were also noted in a study by Kakehi \textit{et al.}~\cite{kakehi_identification_2021}, where the authors used a different object detection model (ResNeXt-101(32 x 8d), part of Detectron2~\cite{wu2019detectron2}) to detect and identify Pacific oyster larvae in images taken from plankton net samples.
The training dataset for this study was generated using a {\it human-in-the-loop} method, where an expert labelled a part of the dataset, which was then used by the object detection model to label further images (which were then expertly verified).
The authors identified problems in the model whereby it could not easily distinguish larvae with obscured outlines, or that overlapped with other objects.
Two possible solutions to the identification problem were proposed: either applying a deep learning method to improve the image quality, or training on a larger dataset that includes more images of larvae with obscured outlines or overlaps.

The issue of object overlap has been addressed by Ruiz-Santaquiteria \textit{et al.}~\cite{RUIZSANTAQUITERIA2020103271}, who applied a Mask R-CNN approach to images of ten species of diatoms.
Mask R-CNN is a development of Faster R-CNN that adds instance segmentation (i.e., it identifies individual instances of objects within regions of interest)~\cite{He2017}.
The authors compared the Mask R-CNN to a SegNet~\cite{SegNet}, which instead performs semantic segmentation (pixels are assigned to a given class, but instances of the same class are not separated).
While SegNet was better on average at detecting true positives (i.e., higher average recall of 0.95, versus 0.86 for Mask R-CNN), Mask R-CNN performed much better in terms of fewer false positives (i.e., higher average specificity of 0.91, versus 0.60 for SegNet), and was better able to separate individual diatoms.

Image quality in general is a major consideration for computer vision problems.
Shi \textit{et al.}~\cite{shi2019study} trained a YOLOv2 model on digital holographic images of zooplankton, of varying sharpness, and noted that a training dataset composed of images with at least an intermediate sharpness assessment index (i.e., at least $0.6$ on a 0-to-1 scale) produced a much better result --- training images with sharpness in the $0.5$ to $0.6$ interval gave a precision of $0.85$ and a recall of $0.70$, whereas the $0.6$ to $0.7$ interval gave a precision of $0.94$ and a recall of $0.89$.

For the purposes of real-world applications of plankton detection, such as real-time monitoring using a remotely operated vehicle, speed of detection is an important consideration.
Li \textit{et al.}~\cite{li_plankton_2021} applied variants of YOLOv3, as well as Faster R-CNN, to the WHOI-Plankton dataset to perform identification and classification \cite{orenstein2015whoi}.
Detection times for YOLOv3 models were two orders of magnitude faster (8 to 51~ms) than the Faster R-CNN (814 to 893~ms), which makes the YOLOv3 models more suitable for real-time applications in this context.
The authors integrated a DenseNet structure into the YOLOv3-dense model.
DenseNet is a structure in which all layers of the network are connected to one another, such that as much information as possible is passed through the network \cite{huang2016}.
Given the small visible differences that can exist between various plankton taxa, this is particularly important.

As the field of deep learning continues to evolve, other more recent methods are being applied to the field of plankton detection, an example of which is shown in Fig.~\ref{figure2}a-f.
In this example, simulated training images bearing similar features to the plankton being studied are generated (Fig.~\ref{figure2}a).
These images are used to train a U-Net model (Fig.~\ref{figure2}b), which outputs segmentation masks corresponding to the positions of the given species (Fig.~\ref{figure2}c).
This trained model is then applied to true experimental images (Fig.~\ref{figure2}d), and corresponding species-specific segmentation masks (Fig.~\ref{figure2}e) are used for cell counting in the experimental image (Fig.~\ref{figure2}f).
Such a workflow is employed in the digital microscopy deep learning framework DeepTrack 2.0 \cite{helgadottir2019digital,midtvedt2021}.
We demonstrate this task using a U-Net architecture, providing training data and analysis code in the associated GitHub repository \cite{bachimanchi_planktonreview_2023}.

With emerging methods, it is also becoming possible to detect objects using only a single, unlabelled image to train a deep learning model.
Such is the case with LodeSTAR \cite{midtvedt_single-shot_2022}, a recently developed method shown to accurately detect multiple instances of the dinoflagellate \textit{Noctiluca scintillans}, after training on a single image, as demonstrated in Fig.~\ref{figure2}g-i.
This is achieved by exploiting an object's symmetry to train a neural network.
In this example, a pair of \textit{N. scintillans} cells (Fig.~\ref{figure2}g) are detected by training the LodeSTAR model (Fig.~\ref{figure2}h) on a training image of a single cell (Fig.~\ref{figure2}i, inset).
Depending on the resolution of the training image, LodeSTAR could detect either a specific feature of the cells (e.g., the tentacle attachment point in Fig.~\ref{figure2}i, left panel) or the entire cells (Fig.~\ref{figure2}i, right panel).
We provide detailed tutorials on how to use the LodeSTAR model for detecting plankton from microscopy videos in our GitHub repository~\cite{bachimanchi_planktonreview_2023}.

{\bf Plankton classification.} 
After detection follows the more complex task of classifying the individual plankton to a taxonomic group.
Several convolutional neural networks have been applied to plankton classification problems over the last decade~\cite{kuang_deep_2015}. Summaries of the studies we mention in this section are given in Table~1.

Even when using relatively large training datasets, additional data may still be needed to train an effective model. This can be achieved through {\it data augmentation}, i.e., the generation of additional training images derived from the original training data.
If a training dataset is too small, it can result in {\it overfitting}, whereby the model interprets noise in the data as a true signal, resulting in a model that has learned to work well with the training dataset, but that fails to generalise to new, unseen data.
In such cases, data augmentation is used to produce more, transformed images, thus bolstering the training dataset.
This augmentation can be achieved through, for example, various transformations of the existing images, such as rotation, cropping, or re-scaling, or through the addition of noise~\cite{kuang_deep_2015,dai_zooplanktonet_2016,correa_deep_2017}.
In their paper on the ZooplanktoNet model, Dai \textit{et al.}~\cite{dai_zooplanktonet_2016} compared the classification accuracy of several popular deep learning models on a zooplankton dataset, with and without data augmentation.
The accuracy increased by an average of 9.7\% when data augmentation was applied (albeit quadrupling the training time).
Likewise, Correa \textit{et al.}~\cite{correa_deep_2017} noted a 17.5\% increase in accuracy after application of data augmentation to a 19-class microalgal dataset.

Regardless of the size of the dataset, there are likely to be classes that are underrepresented, e.g., rare taxa that are detected less often, but whose accurate identification may be vital. For example, this can be the case in the identification of invasive species during the early stages of invasion.
However, rare taxa can have a substantial impact on a model’s performance.
For example, Luo \textit{et al.}~\cite{luo_automated_2018} performed a classification of 75,000 images collected using the \textit{In Situ} Ichthyoplankton Imaging System (\textit{IS}IIS)~\cite{Cowen2008}.
While the average precision achieved was 0.84, exclusion of the twelve rarest taxa from consideration boosted the average precision to 0.907.

As noted by Li \textit{et al.}~\cite{li_plankton_2021}, data augmentation is often applied to the whole training dataset, meaning that if there is a class imbalance, the data augmentation will not address this.
To augment the training data for rare taxa specifically, the authors used a Cycle Generative Adversarial Network ({\it CycleGAN}) model~\cite{Zhu2020}.
A CycleGAN is made up of two pairs of components --- two generators that generate realistic-looking images, and two discriminators, whose purpose is to differentiate between real and generated images.
With each iteration, the generators become better at creating realistic-looking images, while the discriminators become better at detecting them.
Ultimately, the goal is to generate realistic-looking images that look convincingly similar to the originals.
In Li \textit{et al.}, two taxa were randomly selected as ``rare taxa'', and a lower number of true images of these taxa were included in the training dataset (around one-fifth the number of images included for non-rare taxa).
The CycleGAN model was then used to generate images of these two taxa so that the total number of training images was comparable to those of the non-rare taxa.
The result was that the average precision for the two rare taxa augmented with the CycleGAN model increased by 1.93\% (from 0.91 to 0.93) and 6\% (from 0.92 to 0.98), for the randomly-selected classes ``Pennate'' and ``Prorocentrum'', respectively, compared to when these taxa were not augmented.

The issue of class imbalance in plankton datasets has also been addressed using the technique of {\it transfer learning}.
This involves training a model on one dataset (often an unrelated, general dataset, which allows the model to learn broader patterns in image data), and then fine-tuning parts of the model by retraining with data from the problem of interest.
This can also help address the issue of long training times for new models.
In a study on the WHOI-Plankton dataset, Lee \textit{et al.}~\cite{lee2016plankton} initially trained a model on class-normalised data (i.e., any class with more than 5,000 images was randomly subsampled down to 5,000), then fine-tuned on the entire dataset to ensure that population data (the frequency of each class) were not lost.
This method improved the accuracy of the smaller classes while retaining the accuracy of the five most abundant (which composed over 90\% of the training and test data in the full dataset).
Transfer learning has also been applied to plankton studies where class imbalance was not the main focus. 
The ImageNet dataset~\cite{Deng2009} (containing over 14 million images as of March 2023) has been used in multiple transfer learning studies (e.g., Refs.~\cite{orenstein2015whoi}, \cite{orenstein2017transfer}, \cite{macneil_plankton_2021}), resulting in greater accuracy than training on the plankton dataset alone.
In the case of Orenstein and Beijbom's study on multiple plankton datasets~\cite{orenstein2017transfer}, the accuracy increase was around 10\%.

In addition to performance improvement via the aforementioned strategies, improved performance has also been achieved in plankton classification through the inclusion of context metadata (i.e., additional data about the sample beyond just the image itself).
Ellen \textit{et al.}~\cite{ellen_improving_2019} incorporated into their model combinations of three different categories of metadata: Geometric (i.e., containing measurements from the original image before adjustment to be fed into the network), geotemporal (e.g., longitude, latitude, and season), and hydrographic (e.g., chlorophyll \textit{a} (Chl \textit{a}) fluorescence and salinity).
With a baseline accuracy of up to around 0.89 (depending on the dataset), addition of all three categories of metadata gave an accuracy gain of around 1.5\%, up to 0.91.
With further improvements to the network, a final classification accuracy of 0.93 was achieved.
Of the three categories listed above, it was noted that geotemporal and hydrographic metadata were the most informative, with addition of geometric metadata alone having minimal impact on prediction accuracy.

Detection and classification of plankton are challenging tasks, but with the continued developments in the field of deep learning, as illustrated by the above examples, we believe that such methods will continue to speed up and automate these processes.

\begin{figure*}[pt]
    \centering
    \includegraphics[scale = 0.85, angle=0]{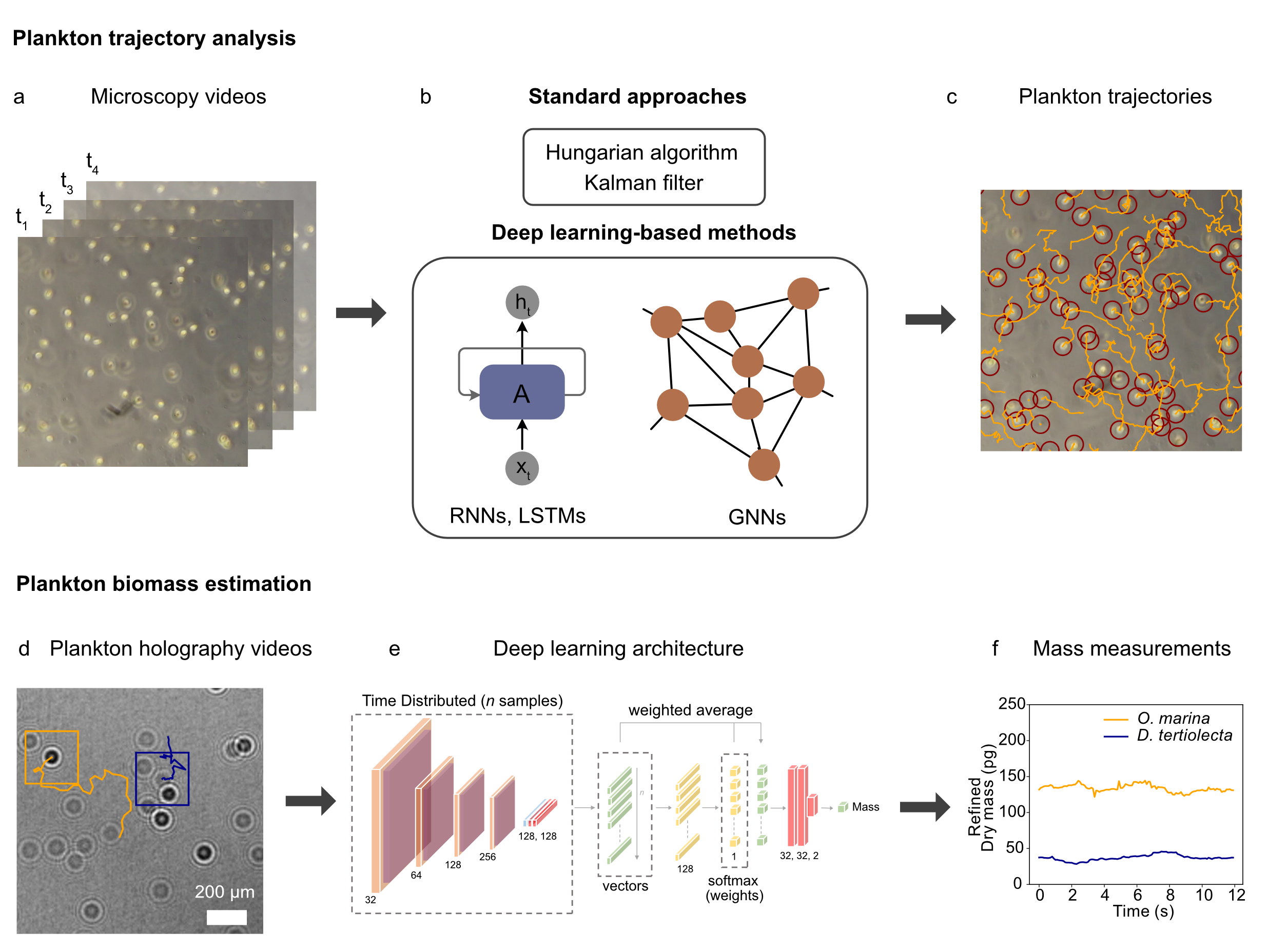}
    \caption{
    \label{figure3}
    {\bf Quantitative analysis of plankton behaviours.}
    {\bf a}-{\bf c} Plankton trajectory analysis.
    {\bf a} Snapshots of plankton species {\it Oxyrrhis marina} taken with a bright-field microscope at $10\times$ magnification are analysed to extract the plankton trajectories.
    {\bf b} Techniques for trajectory linking. Standard approaches (top) include the Hungarian algorithm and linear sum assignment. Deep learning approaches (bottom) include recurrent neural networks (RNNs), long short-term memory networks (LSTMs) (left), and graph neural networks (GNNs) (right).
    {\bf c} Reconstructed plankton trajectories.
    {\bf d}-{\bf f} Plankton biomass estimation.
    The biomass of plankton cells can be estimated by imaging them under quantitative phase microscopy methods such as holography~\cite{bachimanchi2022microplankton}
    {\bf d} Plankton species {\it O. marina} (orange) and {\it D. tertiolecta} (blue) are tracked under a holographic microscope.
    {\bf e} Image crops of size 64 px $\times$ 64 px along the trajectories of respective species are analysed through a deep learning model called WAC-Net~\cite{bachimanchi2022microplankton}.
    {\bf f} Dry mass values (in picograms) of an exemplar of {\it O. marina} (orange) and one of {\it D. tertiolecta} (blue) over time.
    Figures {\bf d}-{\bf f} are adapted from Ref.~\cite{bachimanchi2022microplankton}, with some of the authors of this work being part of the cited reference.
    }

\end{figure*}
\section{Plankton foraging and migration}

\noindent Motility is a fundamental characteristic of living organisms, enabling their interaction with the surrounding environment~\cite{kiorboe2009mechanistic}.
The internal cellular processes that govern an organism's motion have evolved to perform specific major tasks such as reproduction, foraging, migration, defence, and communication. 
For instance, terrestrial organisms such as zebras follow waves of freshly grown vegetation after winter months for active grazing. 
Similarly, planktonic organisms often perform diurnal vertical migration where organisms move to deeper levels of the ocean during the day to avoid predators, and come back to the uppermost layers during the night in search of food and mating opportunities.
More prominently, zooplankton shelter in deeper, darker strata during the daytime to avoid visual predators, while flagellated phytoplankton sometimes do the opposite, collecting nutrients at depth during the night and surfacing to harvest light during the day \cite{pinti2023global}.
The behaviours associated with foraging and migration are easy to observe in larger animals, such as terrestrial ungulates and larger aquatic organisms, which can be followed by acoustic methods \cite{bandara2021two}.
In contrast, the behaviours of microscopic plankton are mainly inferred through video-microscopy-based methods.
However, experiments with these smaller organisms are performed under simulated conditions due to limitations in microscopic system design~\cite{krishnamurthy_scale-free_2020}, although the high-quality video imaging data still allows for successful extraction of individual cell information.
In the previous section, we discussed deep-learning-based image analysis methods for detection and identification of plankton cells from microscopy videos.
This section will explore how the positions of the detected plankton in multiple frames can be linked to determine swimming trajectories, and how we can extract the statistical properties of such trajectories.
Furthermore, we will discuss how {\it reinforcement learning} can be used to model plankton swimming behaviours in various simulated systems.
    
{\bf Deep-learning-based trajectory analysis.}
The study of microplankton motility is essential to understand how swimming patterns mediate foraging behaviours, and which optimisation strategies can be used to improve prey encounter rates while avoiding predators~\cite{visser_plankton_2006}.
In this view, different mathematical models have been developed to categorise the motions of various organisms.
A microscopic particle whose motion is solely governed by the thermal collisions with neighbouring liquid molecules performs Brownian motion, the simplest form of random motion. Non-motile viruses and bacteria (when suspended in liquid) belong to this category.
Mathematically, the motion of a microscopic organism is characterised by the space explored by the organism over time. This can be be inferred by measuring the relationship between the mean square displacement (MSD) of the particle with time ($t$), which grows linearly (${\rm MSD}\propto t$) for particles and organisms following Brownian motion \cite{munoz2021objective}.
The proportionality factor which links the MSD and $t$ is know as the diffusion coefficient ($D$) of the particle trajectory (${\rm MSD} = 2Dt$).
However, a majority of organisms are known to have an active motion mechanism, which results in a non-linear relationship between MSD and time (${\rm MSD} \propto~t^\alpha$). This phenomenon is generally referred to as ``anomalous diffusion". 
Due to a widespread occurrence of anomalous diffusion in the behaviour of various organisms, a theoretical framework has been developed to further characterise their motions based on the spatial and temporal correlations of trajectories.
This has led to the development of several mathematical models, such as Lévy walks, fractional Brownian motion, continuous time random walk and scaled-Brownian motion.
Various methods to determine (anomalous) diffusion properties have been recently evaluated within the ``Anomalous Diffusion Challenge'', concluding that deep-learning-based methods outperform more classical approaches \cite{munoz2021objective}.
    
Heterotrophic dinoflagellates, such as {\it Oxyrrhis marina}, perform helical Lévy walks to optimise random searching in environments with low prey densities~\cite{bartumeus_helical_2003}.
At a larger scale, millimeter-sized organisms, such as copepods, diffuse by multi-fractal random walks to optimise foraging~\cite{schmitt_multifractal_2001}.

The majority of studies employ conventional object-association approaches, such as the Hungarian algorithm and Kalman filtering, to identify moving particles across consecutive frames of a video. These methods establish links between particles based on their proximity to previous positions, and more advanced algorithms also take into account the direction and speed from the previous timestep to improve linking, e.g., when particles overlap~\cite{chenouard_objective_2014, ulman_objective_2017, godinez_tracking_2015}.
Although these techniques work well in many experimental cases, they still require manual input of parameters, can be error-prone (for instance when the cells are dense and motion is unpredictable), and are limited in their ability to accurately analyse large amounts of data.
More importantly, these traditional methods often struggle to measure accurate cell trajectories during scenarios such as cell division (parent cell dividing into daughter cells) and feeding events (predator cells consuming prey cells), as the number of cells in close proximity changes during these events.
    
In recent years, deep-learning-based methods hold promise for analysing and modelling plankton trajectories.
Although deep-learning-based trajectory analysis methods have not been extensively used in plankton studies yet, they have the potential to provide a new perspective on the behaviour of these organisms.
We highlight some of the possibilities below, and show an example application (Fig.~3a-c) in our GitHub repository \cite{bachimanchi_planktonreview_2023}.
A class of deep sequence modelling methods such as recurrent neural networks (RNNs), long short-term memory (LSTMs), and transformers, are useful in analysing sequential data ~\cite{spilger_recurrent_2020, spilger_deep_2021, yao_deep-learning_2020, pineda_geometric_2023}.
Transformers, which are built on the attention mechanism in deep learning (explained in more detail in the next paragraph), can effectively learn and infer associations between different time points in a trajectory by focusing on the most relevant information at each step of the computation~\cite{ying_transformers_2021}.
This technique has been applied to various domains, such as computer vision, natural language processing, speech recognition, and self-supervised image classification, with groundbreaking results \cite{vaswani2017attention, dosovitskiy2020image, caron2021emerging}.
In deep sequence-based models as mentioned here, trajectories are considered as a sequence of time points, and the associations between the time points are learned during training.
More recently, advanced geometric deep learning models such as graph neural networks (GNNs) are proving to be useful in extracting valuable information from the trajectories, including hidden patterns and behaviours~\cite{lee_graph_2021, kipf_semi-supervised_2017}.
GNNs are a type of deep learning model designed to operate on graphs, which can capture both spatial and temporal dependencies between graph nodes. 
In the case of trajectories, each linked trajectory can be represented as a graph, where each point along the trajectory represents a node, and a connection between relevant nodes is called an edge.
As a starting point, the GNN is provided with a complete graph, where all the positions of the objects (plankton) in all the frames of the video are considered as nodes, and all the possible edges within a likelihood radius connect these nodes.
The likelihood radius is related to the range each particle can cover within two frames of a video.
The GNN then learns to correctly classify edges that complete a single plankton trajectory, effectively treating the problem as an edge classification task.
The advantage of GNNs is that, as a deep learning method, the network can be trained with species-specific trajectories to capture complex patterns of movement that are difficult to capture using traditional methods.

Lately, attention-based fingerprinting GNNs, which combine the strengths of transformers (as mentioned above) and GNNs, have shown great promise in a wide range of applications~\cite{pineda_geometric_2023}. 
These models can be used not only for trajectory linking, but also for quantifying the motion parameters of individual trajectories without the need for explicit trajectory linking.
This attention-based technique can encode and predict biological events such as cell division, cell merging, and feeding events. 
In most applications, the ultimate aim of trajectory linking is to understand the underlying behaviours of trajectories. 
By treating the task as a node regression problem, this technique can predict the diffusion coefficient of a trajectory without the need for linking. 
Moreover, by treating the task as a graph-level classification problem, attention-based GNNs can predict the underlying diffusion model itself (such as fractional Brownian motion or continuous-time random walk).
This method could prove useful in many plankton applications, including understanding optimal swimming characteristics in predator-prey regimes and identification of changes in the diffusion model over time.
    
{\bf Deep-learning-based search optimisation.}
Apart from trajectory analysis, deep learning has also been used for search optimisation in plankton ecology.
Reinforcement learning is a type of machine learning where an agent (e.g., a fictive plankton) learns how to interact with an environment to achieve a specific goal through trial and error, receiving rewards or punishments along the way.
Studies in this direction have adapted deep reinforcement-learning-based algorithms to investigate how plankton search for food and optimise their search strategies, or navigate their swimming under different environmental conditions such as turbulent flows \cite{qiu_active_2022, gustavsson_finding_2017, colabrese_flow_2017, gunnarson_learning_2021, alageshan_machine_2020, qiu_navigation_2022}. 
    
For instance, many plankton species migrate vertically in the turbulent ocean, using a mechanism called gyrotaxis to swim efficiently with or against gravity. 
While some plankton rely on passive mechanisms for gyrotactic stability, they are often sensitive to turbulence, which can upset their alignment. 
To address this, Qiu \textit{et al.} ~\cite{qiu_active_2022} have suggested an active mechanism for gyrotactic stability, that they attempt to simulate using reinforcement learning.
The proposed mechanism involves a model plankton that swims at a constant speed and actively steers in response to hydrodynamic signals encountered in simulations of a turbulent flow. 
By sensing its settling velocity and steering in response to hydrodynamic signals, the swimmer can maintain accurate alignment upwards, allowing for efficient upward migration in turbulent flows. 
This finding has provided some initial insights on the role of active mechanisms used by plankton to maintain stability in a turbulent ocean.

Although these studies provide valuable insights into the foraging behaviour of planktonic organisms, it is important to note that they are based on simplified, idealised environments that do not fully represent the complex oceanic conditions. Thus, future studies should focus on incorporating more realistic natural factors (e.g., flow regimes, predator-prey interactions) into the models to obtain a more accurate representation of the ecological dynamics of planktonic organisms.

\section{Quantitative Analysis of Plankton Biomass}
Phytoplankton blooms result from the rapid accumulation of plankton biomass in a restricted water body. They can positively impact the ecosystem, as planktivorous organisms benefit from a high density of prey. However, large plankton blooms can be rather harmful to their environment, as they limit the amount of light reaching the seafloor and often cause oxygen depletion where they die and decompose. Furthermore, harmful algal blooms can produce toxins, impacting human health and aquaculture \cite{Bates2018, diaz2019impacts, lenzen2021impacts}. As predicted global changes will affect the intensity and frequency of plankton blooms, being able to automatically analyse their biomass is a valuable tool for assessing, monitoring and evaluating the effect of phytoplankton biomass on the ecosystem.

The biomass of plankton has previously been estimated using mathematical models. Nutrient-Phytoplankton-Zooplankton (N-P-Z) models have commonly been used to model plankton dynamics~\cite{franks2002npz}. These models use the interactions among predator, prey, and nutrient availability to simulate the dynamics in the ecosystem. Such a model was used by Frost ~\cite{Frost_1987}, who combined chemical (e.g., concentration of Chl \textit{a}), biological (e.g., photosynthetic rate) and physical (e.g., layer depth, incident solar radiation) parameters to simulate the increase in phytoplankton biomass in the Pacific Ocean. N-P-Z models remain valuable tools for plankton biomass simulations~\cite{franks2002npz} but are limited by the relevance of the parameters chosen for the modelling. 

Deep-learning-based approaches have been developed over the past decades to estimate plankton biomass in the context of climate and ecosystem modelling, and to monitor harmful algal blooms. Some of these approaches, all using neural network algorithms, will be highlighted in the present section, with the goal of giving an overview of possible applications of deep learning for quantifying plankton biomass. 

Deep learning can be used in plankton biomass analysis to find correlations between biological events and environmental parameters.
Aoki \textit{et al.}~\cite{Aoki_Komatsu_Hwang_1999} analysed the long-term changes in zooplankton biomass off the northeastern coast of Japan, incorporating 39 years of zooplankton data collected by vertical hauls and environmental data gathered by the Japan Meteorological Agency.
They trained several dense neural networks (i.e., multilayer perceptrons) with different numbers of layers with variable number of nodes per layer, and compared their performance. 
A traditional \emph{back-propagation} algorithm as discussed before is used to train the networks.
The most efficient model to generalise the data was a simple one-hidden-layer model with few hidden units. 
The long-term variation in zooplankton biomass could be predicted by generating mapping functions between environmental variables and zooplankton biomass. Nevertheless, the range of their training dataset did not fit their data properly; hence, they highlighted the importance of the range of the training dataset when performing such analyses. 
A similar approach was undertaken by Woodd-Walker \textit{et al.}~\cite{Woodd-Walker_Kingston_Gallienne}, who collected plankton samples and recorded environmental and seasonal parameters (latitude, temperature, salinity, density, total incoming radiation, Chl \textit{a}, and diel time) along a transect in the Atlantic. They used environmental parameters to estimate surface plankton biomass using linear regression and a two-layer feed-forward network. As in Aoki \textit{et al.}~\cite{Aoki_Komatsu_Hwang_1999}, the network was able to predict the changes in biomass, but the range of the training dataset limited the accuracy of biomass estimates.
The optimal neural network showed an average prediction over the test dataset with a determination coefficient \textit{r\(^2\)} of 0.47, lower than the \textit{r\(^2\)} obtained with the training dataset (0.77).

Other learning algorithms have also been used in the context of ecosystem modelling. To analyse the impacts of similar environmental conditions on the biomass of various phytoplankton groups, Pan \textit{et al.}~\cite{pan2020environmental} used a multi-layer feedforward network to explore the effects of glacial meltwater on phytoplankton communities. 
They found that different environmental variables were responsible for the changes in the abundance of different taxonomic groups. 
Chlorophyll \textit{a} concentrations of flagellates, prasinophytes and diatoms were best predicted by H2O deep-learning models, with respective \textit{r\(^2\)} values of 0.91, 0.9 and 0.89. H2O models are multi-layer feedforward artificial neural networks trained with stochastic gradient descent (SGD) using back-propagation. 
Cryptophyte Chl \textit{a} concentration was best predicted by a gradient boosting model (\textit{r\(^2\)} 0.97). These models permitted the identification of the most important variables for phytoplankton abundance, such as temperature for cryptophytes and flagellates, and nitrate availability for diatoms and prasinophytes.

In the study by Pan {\it et al.}~\cite{pan2020environmental}, \textit{in situ} data collection was feasible. However, this may not always be the case, so that remote imaging techniques are often favoured to study plankton biomass. For example, Chase \textit{et al.}~\cite{chase2022plankton} assessed the reliability of satellite imaging to accurately quantify diatom biomass in the western North Atlantic using a two-layer feedforward network to find a relationship between Chl \textit{a}, environmental information (salinity and temperature) and concentration of diatom-bound carbon. Their methodology seems promising for climate modelling, but it remains preliminary, as the average uncertainty of the diatom carbon estimated using the neural network was relatively high (65\%).

Besides satellite imaging, aerial imaging from drones can provide valuable data to monitor planktonic activity. This tool can be used as the foundation of ecosystem modelling studies based on deep learning techniques. Pyo \textit{et al.}~\cite{Pyo_Hong_Jang_Park_Park_Noh_Cho_2022} developed a one-dimensional convolutional neural network, capable of quantitatively and qualitatively assessing algal blooms when applied to hyperspectral images captured by drones. The network predicted the trends of Chl \textit{a}, phycocyanin, lutein, fucoxanthin, and zeaxanthin with respective \textit{r\(^2\)} values of 0.87, 0.71, 0.76, 0.78, and 0.74, in comparison to observed data. The method remains to be perfected, as image borders have distorted information which might bias the interpretation of the signal.  Still, \textit{ex situ} data collected using drone or satellite imaging is a promising tool for studying the composition of algal blooms.
Such data can even inform on the vertical distribution of phytoplankton. 
For example, Sammartino \textit{et al.}~\cite{Sammartino_Marullo_Santoleri_Scardi_2018} used an error back-propagation neural network (BPN) trained using Chl \textit{a} profiles, geographical data, depth and temperature obtained from \textit{in situ} data and satellite observations to study the vertical profile of Chl \textit{a}. The prediction performance of the BPN was assessed using observed data. For both \textit{in situ} and satellite datasets, the  BPN had a fairly good predicting performance with respective \textit{r\(^2\)} values of 0.69 and 0.63. In conjunction with spatial scales, deep learning approaches can also estimate plankton biomass at temporal scales. 
Martinez \textit{et al.}~\cite{Martinez_2020} found that a {\it multi-layer perceptron} (MLP, a neural network with dense layers) performed better than a support vector regression (SVR) approach to reconstruct the long-term phytoplankton variability by estimating surface Chl \textit{a} ($r^2 > 0.6$ vs. $r^2 < 0.5$). Even though this deep learning algorithm underestimated the amount of Chl \textit{a}, this model is still considered a promising tool for monitoring phytoplankton activity.

As phytoplankton blooms often occur at a large scale, photographing these events has been a favoured method to grasp their extent. With the improvement of the reliability and availability of imaging technologies, the image capture of phytoplankton blooms has become easier, and neural networks have become the preferred tool to analyse such datasets.
More complex networks were recently used to quantify the plankton biomass to predict harmful algal blooms. For example, Liu \textit{et al.}~\cite{liu2022algal} used a combination of a long short-term memory network (LSTM) and time-frequency wavelength analysis (WA), a model named WLSTM, to estimate algal blooms. This model performed better than other neural networks and LSTM alone for predicting biomass variations at different time scales. This was particularly noticeable for longer time scales. At the daily and monthly scale, the hybrid network predicted the data with respective \textit{r\(^2\)} values of 0.878 and 0.814, while the performance of the other methods was lower, with $r^2 < 0.63$. The hybrid WLSTM model was even able to predict extreme algal bloom events.

Another LSTM  was used to identify the most reliable Chl \textit{a} proxy for predicting changes in algal biomass. Wenxiang \textit{et al.}~\cite{wenxiang2022optimization} found that the change rate and the relative change rate of Chl \textit{a} were more accurate outputs for predicting the changes in biomass than the absolute concentration of Chl \textit{a}, especially in winter and spring. Indeed, the correlation coefficient indicating the relationship between the forecasted and the observed concentration of Chl \textit{a} exceeds 0.84 for all-year-round predictions when using the change rate and the relative change rate of Chl \textit{a}, while it dropped below 0.6 when using the absolute concentration of Chl in winter and spring. This information can help direct future studies aimed at predicting algal blooms.

At a much smaller scale, deep learning algorithms can also be used to estimate the dry mass of individual plankton cells. Bachimanchi \textit{et al.}~\cite{bachimanchi2022microplankton} used a combination of a regression U-Net (RU-Net) and a weighted-average convolutional neural network (WAC-Net) to analyse holograms of plankton. The RU-Net is a modified convolutional neural network, that segments the image into several heatmaps. This output can be processed by the WAC-Net to predict the dry mass and radius. The dry mass estimations obtained by WAC-Net and by the “volume-to-carbon” method for several species zooplankton and phytoplankton have a correlation coefficient of 0.988. Thanks to this network combination, it becomes feasible to follow the growth of a single cell, something that is impossible with classical methods (Fig.~3d-f).
Through these examples, we can see that deep learning has revolutionised the field of plankton biomass estimation. Even though these networks have different levels of complexity, they have eased the study of plankton at various temporal and spatial scales, thanks to non-invasive imaging techniques, providing endless opportunities to model climate and ecosystems, and monitor harmful algal blooms.  

\section{Opportunities}
    In the following section, we explore the most recent advancements in deep learning in light of their their potential applications in the field of plankton ecology. These developments offer new opportunities to overcome challenges and pave the way for innovative solutions.

    One of the biggest hurdles in many deep-learning applications is the acquisition of labelled data. However, recent advances in unsupervised and self-supervised deep learning algorithms have shown promise in addressing this problem. For instance, CycleGANs, which are part of the generative adversarial network family, can, in addition to their use in generating new data, be utilised for segmentation tasks in plankton images even when there is a limited amount of manually annotated data. 
    Since synthetically-generated masks with similar morphologies can be used as ground truth data for real plankton images, CycleGANs can be trained on unpaired images and ground truth data to produce accurate segmentations.
    
    Moreover, microscopy imaging data of plankton can be synthetically generated through either GANs or by simulating plankton-like objects. This method allows for ground-truth data to be known beforehand and controlled, which can help overcome the challenges associated with manual annotation of data. As an example, the segmentations of plankton species shown in Fig.~2 were obtained by a U-Net model trained on simulated plankton data. This approach also offers cell classification based on morphological properties, in addition to segmentation.
    
    Self-supervised deep learning algorithms, such as vision transformers (ViTs), have also demonstrated promising results in object segmentation and detection tasks. Unlike supervised algorithms, which rely on labelled examples, self-supervised algorithms use transformed versions of input images themselves as labels. DINO (distillation with no labels), a self-supervised vision transformer belonging to the ViT family, was able to accurately segment objects from images without any labelled data. Given the diverse morphologies of plankton, self-supervised vision transformers can be used for segmentation and classification of complex taxa.

    In conclusion, the field of AI-driven microscopy is advancing quickly, offering new opportunities for identifying, segmenting, and tracking individual plankton organisms. These developments are expected to revolutionise our understanding of the lower aquatic food web, providing insights that were previously unattainable. With the help of modern microscopy methods and deep learning algorithms, it will soon be possible to observe large numbers of organisms at high spatial and temporal resolutions, enabling us to better understand their interactions and ecology. These breakthroughs are poised to catalyse further progress in plankton ecology in the years to come.

\section{Guidelines for using deep learning in plankton ecology}

To ensure the effective use of deep learning in plankton ecology research, we propose the following specific guidelines, on top om more general guidelines that can be found for example in Ref.~\cite{cichos2020machine}:

\begin{enumerate}

    \item \textbf{Model selection and development:}
    During experimental design, it is important to choose an appropriate model for the task being performed. For example:
    \begin{itemize}
        \item Regression models: where a value or quantity needs to be predicted from the given data (e.g., dry mass prediction in section III).
        \item Classification models: where the given data (e.g., images of planktons) need to be classified into a particular class (such as species).
        \item Generative models: when the data is insufficient, and additional data is needed to train models.
        \item Image-to-image transformation models: where an image from one modality is converted to a different modality for easier analysis (e.g., a microscopy image to a binary map (segmentation map) for classification and detection purposes (see Fig.~2)).    
    \end{itemize}

    In the GitHub repository accompanying this article, we provide an example model using DeepTrack 2.0.
    Other open-source models are also readily available, and we believe the studies mentioned in this article will help direct readers towards models suitable for addressing their own research questions.

    \item \textbf{Data Preparation and Annotation:} 
    Once the task and appropriate model have been defined, one needs to prepare the input data in the right way.
    For instance, for classification tasks, high quality plankton image data, classified by an expert taxonomist, is required.
    For regression tasks, such as predicting the physical parameters of a plankton (size, aspect ratios, dry mass, etc.), ground truth data can be obtained by standard methods as a representative set.
    Additionally, for segmentation tasks, simulating plankton images (as shown in Fig.~2) has its own advantages, as ground truth masks are already known.
    
    \item \textbf{Training and Validation:} 
    Once the data and the model are prepared, the next step is to train the model to maximise its accuracy and minimise its error rate. The data obtained in the previous step must be divided into training, validation, and test data, each representative of the dataset as a whole. The training is the step where the algorithm learns from the data. During training, the performance of the model should be evaluated regularly using metrics like those described earlier in this review. If required, data augmentation can be applied to improve the model's capacity to generalise across variations in the plankton's appearance. The validation dataset is used to evaluate the performance of the model. The model does not learn from the validation dataset, but the results are used to tune the model parameters. Lastly, the test dataset is used to get the final evaluation of the model's performance.   
    
    \item \textbf{Interpretation and Validation of Results:} 
    Once the trained model has been applied to the data, it is important to interpret the data in the context of plankton ecology. Understanding how the model makes its decisions, and assessing any uncertainties, is vital. At least part of the model's predictions should also be compared to results obtained from traditional methods, to ensure the accuracy and reliability of the model. 
    For example, manual identification of species by taxonomists in the classifications tasks, and estimation of biomass by elemental analysis methods for the biomass estimation tasks.
    
    \item \textbf{Reproducibility and Open Science:} 
    Given the ongoing reproducibility crisis across scientific fields \cite{Baker2016}, we strongly recommend the sharing of analysis code, models, and datasets used in studies, along with detailed documentation on their use. Platforms such as GitHub can be used for this purpose, and sharing data in this way facilitates collaboration and scientific progress.
    
\end{enumerate}

By adhering to these guidelines, researchers can ensure the effective and responsible use of deep learning in plankton ecology, leading to robust and insightful outcomes in the field. The guidelines highlight the importance of proper data preparation, model selection, training and validation, result interpretation, and promoting reproducibility and open science.

\bibliography{references}

\providecommand{\noopsort}[1]{}\providecommand{\singleletter}[1]{#1}%
\begin{thebibliography}{92}%
\makeatletter
\providecommand \@ifxundefined [1]{%
 \@ifx{#1\undefined}
}%
\providecommand \@ifnum [1]{%
 \ifnum #1\expandafter \@firstoftwo
 \else \expandafter \@secondoftwo
 \fi
}%
\providecommand \@ifx [1]{%
 \ifx #1\expandafter \@firstoftwo
 \else \expandafter \@secondoftwo
 \fi
}%
\providecommand \natexlab [1]{#1}%
\providecommand \enquote  [1]{``#1''}%
\providecommand \bibnamefont  [1]{#1}%
\providecommand \bibfnamefont [1]{#1}%
\providecommand \citenamefont [1]{#1}%
\providecommand \href@noop [0]{\@secondoftwo}%
\providecommand \href [0]{\begingroup \@sanitize@url \@href}%
\providecommand \@href[1]{\@@startlink{#1}\@@href}%
\providecommand \@@href[1]{\endgroup#1\@@endlink}%
\providecommand \@sanitize@url [0]{\catcode `\\12\catcode `\$12\catcode `\&12\catcode `\#12\catcode `\^12\catcode `\_12\catcode `\%12\relax}%
\providecommand \@@startlink[1]{}%
\providecommand \@@endlink[0]{}%
\providecommand \url  [0]{\begingroup\@sanitize@url \@url }%
\providecommand \@url [1]{\endgroup\@href {#1}{\urlprefix }}%
\providecommand \urlprefix  [0]{URL }%
\providecommand \Eprint [0]{\href }%
\providecommand \doibase [0]{https://doi.org/}%
\providecommand \selectlanguage [0]{\@gobble}%
\providecommand \bibinfo  [0]{\@secondoftwo}%
\providecommand \bibfield  [0]{\@secondoftwo}%
\providecommand \translation [1]{[#1]}%
\providecommand \BibitemOpen [0]{}%
\providecommand \bibitemStop [0]{}%
\providecommand \bibitemNoStop [0]{.\EOS\space}%
\providecommand \EOS [0]{\spacefactor3000\relax}%
\providecommand \BibitemShut  [1]{\csname bibitem#1\endcsname}%
\let\auto@bib@innerbib\@empty
\bibitem [{\citenamefont {Bachimanchi}\ \emph {et~al.}(2023)\citenamefont {Bachimanchi}, \citenamefont {Selander},\ and\ \citenamefont {Volpe}}]{bachimanchi_planktonreview_2023}%
  \BibitemOpen
  \bibfield  {author} {\bibinfo {author} {\bibfnamefont {H.}~\bibnamefont {Bachimanchi}}, \bibinfo {author} {\bibfnamefont {E.}~\bibnamefont {Selander}},\ and\ \bibinfo {author} {\bibfnamefont {G.}~\bibnamefont {Volpe}},\ }\href@noop {} {\bibinfo {title} {Deep-learning-in-plankton-ecology}},\ \bibinfo {howpublished} {\url{https://github.com/softmatterlab/Deep-learning-in-plankton-ecology/}} (\bibinfo {year} {2023})\BibitemShut {NoStop}%
\bibitem [{\citenamefont {Racault}\ \emph {et~al.}(2014)\citenamefont {Racault}, \citenamefont {Platt}, \citenamefont {Sathyendranath}, \citenamefont {Ağirbaş}, \citenamefont {{Martinez Vicente}},\ and\ \citenamefont {Brewin}}]{Racault2014}%
  \BibitemOpen
  \bibfield  {author} {\bibinfo {author} {\bibfnamefont {M.-F.}\ \bibnamefont {Racault}}, \bibinfo {author} {\bibfnamefont {T.}~\bibnamefont {Platt}}, \bibinfo {author} {\bibfnamefont {S.}~\bibnamefont {Sathyendranath}}, \bibinfo {author} {\bibfnamefont {E.}~\bibnamefont {Ağirbaş}}, \bibinfo {author} {\bibfnamefont {V.}~\bibnamefont {{Martinez Vicente}}},\ and\ \bibinfo {author} {\bibfnamefont {R.}~\bibnamefont {Brewin}},\ }\bibfield  {title} {\bibinfo {title} {{Plankton indicators and ocean observing systems: support to the marine ecosystem state assessment}},\ }\href {https://doi.org/10.1093/plankt/fbu016} {\bibfield  {journal} {\bibinfo  {journal} {Journal of Plankton Research}\ }\textbf {\bibinfo {volume} {36}},\ \bibinfo {pages} {621} (\bibinfo {year} {2014})}\BibitemShut {NoStop}%
\bibitem [{\citenamefont {Hensen}(1887)}]{Hensen1887}%
  \BibitemOpen
  \bibfield  {author} {\bibinfo {author} {\bibfnamefont {V.}~\bibnamefont {Hensen}},\ }\bibfield  {title} {\bibinfo {title} {{Kapitel 1: {\"{U}}ber die Bestimmung des Plankton's oder des im Meere treibenden Materials an Pflanzen und Thieren}},\ }\href {https://oceanrep.geomar.de/id/eprint/58770/} {\bibfield  {journal} {\bibinfo  {journal} {Jahresbericht der Commission zur Wissenschaftlichen Untersuchung der Deutschen Meere in Kiel : f{\"{u}}r die Jahre ...}\ }\textbf {\bibinfo {volume} {12-16}},\ \bibinfo {pages} {1} (\bibinfo {year} {1887})}\BibitemShut {NoStop}%
\bibitem [{\citenamefont {Butterwick}\ \emph {et~al.}(1982)\citenamefont {Butterwick}, \citenamefont {Heaney},\ and\ \citenamefont {Talling}}]{Butterwick1982}%
  \BibitemOpen
  \bibfield  {author} {\bibinfo {author} {\bibfnamefont {C.}~\bibnamefont {Butterwick}}, \bibinfo {author} {\bibfnamefont {S.~I.}\ \bibnamefont {Heaney}},\ and\ \bibinfo {author} {\bibfnamefont {J.~F.}\ \bibnamefont {Talling}},\ }\bibfield  {title} {\bibinfo {title} {{A comparison of eight methods for estimating the biomass and growth of planktonic algae}},\ }\href {https://doi.org/10.1080/00071618200650091} {\bibfield  {journal} {\bibinfo  {journal} {British Phycological Journal}\ }\textbf {\bibinfo {volume} {17}},\ \bibinfo {pages} {69} (\bibinfo {year} {1982})}\BibitemShut {NoStop}%
\bibitem [{\citenamefont {Ki{\o}rboe}(2009)}]{kiorboe2009mechanistic}%
  \BibitemOpen
  \bibfield  {author} {\bibinfo {author} {\bibfnamefont {T.}~\bibnamefont {Ki{\o}rboe}},\ }\bibfield  {title} {\bibinfo {title} {A mechanistic approach to plankton ecology},\ }\href {https://doi.org/10.4319/lol.2009.tkiorboe.2} {\bibfield  {journal} {\bibinfo  {journal} {ASLO Web Lectures}\ }\textbf {\bibinfo {volume} {1}},\ \bibinfo {pages} {1} (\bibinfo {year} {2009})}\BibitemShut {NoStop}%
\bibitem [{\citenamefont {Orenstein}\ \emph {et~al.}(2020)\citenamefont {Orenstein}, \citenamefont {Ratelle}, \citenamefont {Brise{\~n}o-Avena}, \citenamefont {Carter}, \citenamefont {Franks}, \citenamefont {Jaffe},\ and\ \citenamefont {Roberts}}]{orenstein2020scripps}%
  \BibitemOpen
  \bibfield  {author} {\bibinfo {author} {\bibfnamefont {E.~C.}\ \bibnamefont {Orenstein}}, \bibinfo {author} {\bibfnamefont {D.}~\bibnamefont {Ratelle}}, \bibinfo {author} {\bibfnamefont {C.}~\bibnamefont {Brise{\~n}o-Avena}}, \bibinfo {author} {\bibfnamefont {M.~L.}\ \bibnamefont {Carter}}, \bibinfo {author} {\bibfnamefont {P.~J.}\ \bibnamefont {Franks}}, \bibinfo {author} {\bibfnamefont {J.~S.}\ \bibnamefont {Jaffe}},\ and\ \bibinfo {author} {\bibfnamefont {P.~L.}\ \bibnamefont {Roberts}},\ }\bibfield  {title} {\bibinfo {title} {The scripps plankton camera system: A framework and platform for in situ microscopy},\ }\href {https://doi.org/10.1002/lom3.10394} {\bibfield  {journal} {\bibinfo  {journal} {Limnology and Oceanography: Methods}\ }\textbf {\bibinfo {volume} {18}},\ \bibinfo {pages} {681} (\bibinfo {year} {2020})}\BibitemShut {NoStop}%
\bibitem [{\citenamefont {Bachimanchi}\ \emph {et~al.}(2022)\citenamefont {Bachimanchi}, \citenamefont {Midtvedt}, \citenamefont {Midtvedt}, \citenamefont {Selander},\ and\ \citenamefont {Volpe}}]{bachimanchi2022microplankton}%
  \BibitemOpen
  \bibfield  {author} {\bibinfo {author} {\bibfnamefont {H.}~\bibnamefont {Bachimanchi}}, \bibinfo {author} {\bibfnamefont {B.}~\bibnamefont {Midtvedt}}, \bibinfo {author} {\bibfnamefont {D.}~\bibnamefont {Midtvedt}}, \bibinfo {author} {\bibfnamefont {E.}~\bibnamefont {Selander}},\ and\ \bibinfo {author} {\bibfnamefont {G.}~\bibnamefont {Volpe}},\ }\bibfield  {title} {\bibinfo {title} {Microplankton life histories revealed by holographic microscopy and deep learning},\ }\href {https://doi.org/10.7554/eLife.79760} {\bibfield  {journal} {\bibinfo  {journal} {eLife}\ }\textbf {\bibinfo {volume} {11}},\ \bibinfo {pages} {e79760} (\bibinfo {year} {2022})}\BibitemShut {NoStop}%
\bibitem [{\citenamefont {Falkowski}(1994)}]{Falkowski1994}%
  \BibitemOpen
  \bibfield  {author} {\bibinfo {author} {\bibfnamefont {P.~G.}\ \bibnamefont {Falkowski}},\ }\bibfield  {title} {\bibinfo {title} {{The role of phytoplankton photosynthesis in global biogeochemical cycles}},\ }\href {https://doi.org/10.1007/BF00014586} {\bibfield  {journal} {\bibinfo  {journal} {Photosynthesis Research}\ }\textbf {\bibinfo {volume} {39}},\ \bibinfo {pages} {235} (\bibinfo {year} {1994})}\BibitemShut {NoStop}%
\bibitem [{\citenamefont {Bates}\ \emph {et~al.}(2018)\citenamefont {Bates}, \citenamefont {Hubbard}, \citenamefont {Lundholm}, \citenamefont {Montresor},\ and\ \citenamefont {Leaw}}]{Bates2018}%
  \BibitemOpen
  \bibfield  {author} {\bibinfo {author} {\bibfnamefont {S.~S.}\ \bibnamefont {Bates}}, \bibinfo {author} {\bibfnamefont {K.~A.}\ \bibnamefont {Hubbard}}, \bibinfo {author} {\bibfnamefont {N.}~\bibnamefont {Lundholm}}, \bibinfo {author} {\bibfnamefont {M.}~\bibnamefont {Montresor}},\ and\ \bibinfo {author} {\bibfnamefont {C.~P.}\ \bibnamefont {Leaw}},\ }\bibfield  {title} {\bibinfo {title} {{Pseudo-nitzschia, Nitzschia, and domoic acid: New research since 2011}},\ }\href {https://doi.org/10.1016/j.hal.2018.06.001} {\bibfield  {journal} {\bibinfo  {journal} {Harmful Algae}\ }\textbf {\bibinfo {volume} {79}},\ \bibinfo {pages} {3} (\bibinfo {year} {2018})}\BibitemShut {NoStop}%
\bibitem [{\citenamefont {HELCOM}(2021)}]{HELCOM}%
  \BibitemOpen
  \bibfield  {author} {\bibinfo {author} {\bibnamefont {HELCOM}},\ }\href@noop {} {\bibinfo {title} {Guidelines for monitoring of phytoplankton species composition, abundance and biomass (2021)}},\ \bibinfo {howpublished} {https://helcom.fi/wp-content/uploads/2020/01/HELCOM-Guidelines-for-monitoring-of-phytoplankton-species-composition-abundance-and-biomass.pdf} (\bibinfo {year} {2021})\BibitemShut {NoStop}%
\bibitem [{\citenamefont {Mackay}\ \emph {et~al.}(2003)\citenamefont {Mackay}, \citenamefont {Jones},\ and\ \citenamefont {Battarbee}}]{mackay2003approaches}%
  \BibitemOpen
  \bibfield  {author} {\bibinfo {author} {\bibfnamefont {A.~W.}\ \bibnamefont {Mackay}}, \bibinfo {author} {\bibfnamefont {V.~J.}\ \bibnamefont {Jones}},\ and\ \bibinfo {author} {\bibfnamefont {R.~W.}\ \bibnamefont {Battarbee}},\ }\href {https://www.geog.ucl.ac.uk/people/academic-staff/anson-mackay/files/20Mackay20ppp.pdf} {\bibinfo {title} {Global change in the holocene}} (\bibinfo {year} {2003})\BibitemShut {NoStop}%
\bibitem [{\citenamefont {Malviya}\ \emph {et~al.}(2016)\citenamefont {Malviya}, \citenamefont {Scalco}, \citenamefont {Audic}, \citenamefont {Vincent}, \citenamefont {Veluchamy}, \citenamefont {Poulain}, \citenamefont {Wincker}, \citenamefont {Iudicone}, \citenamefont {de~Vargas}, \citenamefont {Bittner} \emph {et~al.}}]{malviya2016insights}%
  \BibitemOpen
  \bibfield  {author} {\bibinfo {author} {\bibfnamefont {S.}~\bibnamefont {Malviya}}, \bibinfo {author} {\bibfnamefont {E.}~\bibnamefont {Scalco}}, \bibinfo {author} {\bibfnamefont {S.}~\bibnamefont {Audic}}, \bibinfo {author} {\bibfnamefont {F.}~\bibnamefont {Vincent}}, \bibinfo {author} {\bibfnamefont {A.}~\bibnamefont {Veluchamy}}, \bibinfo {author} {\bibfnamefont {J.}~\bibnamefont {Poulain}}, \bibinfo {author} {\bibfnamefont {P.}~\bibnamefont {Wincker}}, \bibinfo {author} {\bibfnamefont {D.}~\bibnamefont {Iudicone}}, \bibinfo {author} {\bibfnamefont {C.}~\bibnamefont {de~Vargas}}, \bibinfo {author} {\bibfnamefont {L.}~\bibnamefont {Bittner}}, \emph {et~al.},\ }\bibfield  {title} {\bibinfo {title} {Insights into global diatom distribution and diversity in the world’s ocean},\ }\href {https://doi.org/10.1073/pnas.1509523113} {\bibfield  {journal} {\bibinfo  {journal} {Proceedings of the National Academy of Sciences}\ }\textbf {\bibinfo {volume} {113}},\ \bibinfo {pages} {E1516} (\bibinfo {year} {2016})}\BibitemShut {NoStop}%
\bibitem [{\citenamefont {Bucklin}\ \emph {et~al.}(2016)\citenamefont {Bucklin}, \citenamefont {Lindeque}, \citenamefont {Rodriguez-Ezpeleta}, \citenamefont {Albaina},\ and\ \citenamefont {Lehtiniemi}}]{bucklin2016metabarcoding}%
  \BibitemOpen
  \bibfield  {author} {\bibinfo {author} {\bibfnamefont {A.}~\bibnamefont {Bucklin}}, \bibinfo {author} {\bibfnamefont {P.~K.}\ \bibnamefont {Lindeque}}, \bibinfo {author} {\bibfnamefont {N.}~\bibnamefont {Rodriguez-Ezpeleta}}, \bibinfo {author} {\bibfnamefont {A.}~\bibnamefont {Albaina}},\ and\ \bibinfo {author} {\bibfnamefont {M.}~\bibnamefont {Lehtiniemi}},\ }\bibfield  {title} {\bibinfo {title} {Metabarcoding of marine zooplankton: prospects, progress and pitfalls},\ }\href {https://doi.org/10.1093/plankt/fbw023} {\bibfield  {journal} {\bibinfo  {journal} {Journal of Plankton Research}\ }\textbf {\bibinfo {volume} {38}},\ \bibinfo {pages} {393} (\bibinfo {year} {2016})}\BibitemShut {NoStop}%
\bibitem [{\citenamefont {Ruppert}\ \emph {et~al.}(2019)\citenamefont {Ruppert}, \citenamefont {Kline},\ and\ \citenamefont {Rahman}}]{ruppert2019past}%
  \BibitemOpen
  \bibfield  {author} {\bibinfo {author} {\bibfnamefont {K.~M.}\ \bibnamefont {Ruppert}}, \bibinfo {author} {\bibfnamefont {R.~J.}\ \bibnamefont {Kline}},\ and\ \bibinfo {author} {\bibfnamefont {M.~S.}\ \bibnamefont {Rahman}},\ }\bibfield  {title} {{\selectlanguage {english}\bibinfo {title} {Past, present, and future perspectives of environmental {DNA} ({eDNA}) metabarcoding: {A} systematic review in methods, monitoring, and applications of global {eDNA}}},\ }\href {https://doi.org/10.1016/j.gecco.2019.e00547} {\bibfield  {journal} {\bibinfo  {journal} {Global Ecology and Conservation}\ }\textbf {\bibinfo {volume} {17}},\ \bibinfo {pages} {e00547} (\bibinfo {year} {2019})}\BibitemShut {NoStop}%
\bibitem [{\citenamefont {LeCun}\ \emph {et~al.}(2015)\citenamefont {LeCun}, \citenamefont {Bengio},\ and\ \citenamefont {Hinton}}]{LeCun2015}%
  \BibitemOpen
  \bibfield  {author} {\bibinfo {author} {\bibfnamefont {Y.}~\bibnamefont {LeCun}}, \bibinfo {author} {\bibfnamefont {Y.}~\bibnamefont {Bengio}},\ and\ \bibinfo {author} {\bibfnamefont {G.}~\bibnamefont {Hinton}},\ }\bibfield  {title} {\bibinfo {title} {{Deep learning}},\ }\href {https://doi.org/10.1038/nature14539} {\bibfield  {journal} {\bibinfo  {journal} {Nature}\ }\textbf {\bibinfo {volume} {521}},\ \bibinfo {pages} {436} (\bibinfo {year} {2015})}\BibitemShut {NoStop}%
\bibitem [{\citenamefont {Young}\ \emph {et~al.}(2018)\citenamefont {Young}, \citenamefont {Hazarika}, \citenamefont {Poria},\ and\ \citenamefont {Cambria}}]{Young2018}%
  \BibitemOpen
  \bibfield  {author} {\bibinfo {author} {\bibfnamefont {T.}~\bibnamefont {Young}}, \bibinfo {author} {\bibfnamefont {D.}~\bibnamefont {Hazarika}}, \bibinfo {author} {\bibfnamefont {S.}~\bibnamefont {Poria}},\ and\ \bibinfo {author} {\bibfnamefont {E.}~\bibnamefont {Cambria}},\ }\bibfield  {title} {\bibinfo {title} {{Recent Trends in Deep Learning Based Natural Language Processing [Review Article]}},\ }\href {https://doi.org/10.1109/MCI.2018.2840738} {\bibfield  {journal} {\bibinfo  {journal} {IEEE Computational Intelligence Magazine}\ }\textbf {\bibinfo {volume} {13}},\ \bibinfo {pages} {55} (\bibinfo {year} {2018})}\BibitemShut {NoStop}%
\bibitem [{\citenamefont {Chai}\ \emph {et~al.}(2021)\citenamefont {Chai}, \citenamefont {Zeng}, \citenamefont {Li},\ and\ \citenamefont {Ngai}}]{Chai2021}%
  \BibitemOpen
  \bibfield  {author} {\bibinfo {author} {\bibfnamefont {J.}~\bibnamefont {Chai}}, \bibinfo {author} {\bibfnamefont {H.}~\bibnamefont {Zeng}}, \bibinfo {author} {\bibfnamefont {A.}~\bibnamefont {Li}},\ and\ \bibinfo {author} {\bibfnamefont {E.~W.~T.}\ \bibnamefont {Ngai}},\ }\bibfield  {title} {\bibinfo {title} {{Deep learning in computer vision: A critical review of emerging techniques and application scenarios}},\ }\href {https://doi.org/10.1016/j.mlwa.2021.100134} {\bibfield  {journal} {\bibinfo  {journal} {Machine Learning with Applications}\ }\textbf {\bibinfo {volume} {6}},\ \bibinfo {pages} {100134} (\bibinfo {year} {2021})}\BibitemShut {NoStop}%
\bibitem [{\citenamefont {Alzubaidi}\ \emph {et~al.}(2021)\citenamefont {Alzubaidi}, \citenamefont {Zhang}, \citenamefont {Humaidi}, \citenamefont {Al-Dujaili}, \citenamefont {Duan}, \citenamefont {Al-Shamma}, \citenamefont {Santamar{\'{i}}a}, \citenamefont {Fadhel}, \citenamefont {Al-Amidie},\ and\ \citenamefont {Farhan}}]{Alzubaidi2021}%
  \BibitemOpen
  \bibfield  {author} {\bibinfo {author} {\bibfnamefont {L.}~\bibnamefont {Alzubaidi}}, \bibinfo {author} {\bibfnamefont {J.}~\bibnamefont {Zhang}}, \bibinfo {author} {\bibfnamefont {A.~J.}\ \bibnamefont {Humaidi}}, \bibinfo {author} {\bibfnamefont {A.}~\bibnamefont {Al-Dujaili}}, \bibinfo {author} {\bibfnamefont {Y.}~\bibnamefont {Duan}}, \bibinfo {author} {\bibfnamefont {O.}~\bibnamefont {Al-Shamma}}, \bibinfo {author} {\bibfnamefont {J.}~\bibnamefont {Santamar{\'{i}}a}}, \bibinfo {author} {\bibfnamefont {M.~A.}\ \bibnamefont {Fadhel}}, \bibinfo {author} {\bibfnamefont {M.}~\bibnamefont {Al-Amidie}},\ and\ \bibinfo {author} {\bibfnamefont {L.}~\bibnamefont {Farhan}},\ }\bibfield  {title} {\bibinfo {title} {{Review of deep learning: concepts, CNN architectures, challenges, applications, future directions}},\ }\href {https://doi.org/10.1186/s40537-021-00444-8} {\bibfield  {journal} {\bibinfo  {journal} {Journal of Big Data}\ }\textbf {\bibinfo {volume} {8}},\ \bibinfo {pages} {53} (\bibinfo {year} {2021})}\BibitemShut {NoStop}%
\bibitem [{\citenamefont {Abad}\ \emph {et~al.}(2016)\citenamefont {Abad}, \citenamefont {Albaina}, \citenamefont {Aguirre}, \citenamefont {Laza-Mart{\'{i}}nez}, \citenamefont {Uriarte}, \citenamefont {Iriarte}, \citenamefont {Villate},\ and\ \citenamefont {Estonba}}]{Abad2016}%
  \BibitemOpen
  \bibfield  {author} {\bibinfo {author} {\bibfnamefont {D.}~\bibnamefont {Abad}}, \bibinfo {author} {\bibfnamefont {A.}~\bibnamefont {Albaina}}, \bibinfo {author} {\bibfnamefont {M.}~\bibnamefont {Aguirre}}, \bibinfo {author} {\bibfnamefont {A.}~\bibnamefont {Laza-Mart{\'{i}}nez}}, \bibinfo {author} {\bibfnamefont {I.}~\bibnamefont {Uriarte}}, \bibinfo {author} {\bibfnamefont {A.}~\bibnamefont {Iriarte}}, \bibinfo {author} {\bibfnamefont {F.}~\bibnamefont {Villate}},\ and\ \bibinfo {author} {\bibfnamefont {A.}~\bibnamefont {Estonba}},\ }\bibfield  {title} {\bibinfo {title} {{Is metabarcoding suitable for estuarine plankton monitoring? A comparative study with microscopy}},\ }\href {https://doi.org/10.1007/s00227-016-2920-0} {\bibfield  {journal} {\bibinfo  {journal} {Marine Biology}\ }\textbf {\bibinfo {volume} {163}},\ \bibinfo {pages} {149} (\bibinfo {year} {2016})}\BibitemShut {NoStop}%
\bibitem [{\citenamefont {H{\o}ye}\ \emph {et~al.}(2021)\citenamefont {H{\o}ye}, \citenamefont {{\"A}rje}, \citenamefont {Bjerge}, \citenamefont {Hansen}, \citenamefont {Iosifidis}, \citenamefont {Leese}, \citenamefont {Mann}, \citenamefont {Meissner}, \citenamefont {Melvad},\ and\ \citenamefont {Raitoharju}}]{hoye2021deep}%
  \BibitemOpen
  \bibfield  {author} {\bibinfo {author} {\bibfnamefont {T.~T.}\ \bibnamefont {H{\o}ye}}, \bibinfo {author} {\bibfnamefont {J.}~\bibnamefont {{\"A}rje}}, \bibinfo {author} {\bibfnamefont {K.}~\bibnamefont {Bjerge}}, \bibinfo {author} {\bibfnamefont {O.~L.}\ \bibnamefont {Hansen}}, \bibinfo {author} {\bibfnamefont {A.}~\bibnamefont {Iosifidis}}, \bibinfo {author} {\bibfnamefont {F.}~\bibnamefont {Leese}}, \bibinfo {author} {\bibfnamefont {H.~M.}\ \bibnamefont {Mann}}, \bibinfo {author} {\bibfnamefont {K.}~\bibnamefont {Meissner}}, \bibinfo {author} {\bibfnamefont {C.}~\bibnamefont {Melvad}},\ and\ \bibinfo {author} {\bibfnamefont {J.}~\bibnamefont {Raitoharju}},\ }\bibfield  {title} {\bibinfo {title} {Deep learning and computer vision will transform entomology},\ }\href@noop {} {\bibfield  {journal} {\bibinfo  {journal} {Proceedings of the National Academy of Sciences}\ }\textbf {\bibinfo {volume} {118}},\ \bibinfo {pages} {e2002545117} (\bibinfo {year} {2021})}\BibitemShut {NoStop}%
\bibitem [{\citenamefont {Orenstein}\ \emph {et~al.}(2015)\citenamefont {Orenstein}, \citenamefont {Beijbom}, \citenamefont {Peacock},\ and\ \citenamefont {Sosik}}]{orenstein2015whoi}%
  \BibitemOpen
  \bibfield  {author} {\bibinfo {author} {\bibfnamefont {E.~C.}\ \bibnamefont {Orenstein}}, \bibinfo {author} {\bibfnamefont {O.}~\bibnamefont {Beijbom}}, \bibinfo {author} {\bibfnamefont {E.~E.}\ \bibnamefont {Peacock}},\ and\ \bibinfo {author} {\bibfnamefont {H.~M.}\ \bibnamefont {Sosik}},\ }\bibfield  {title} {\bibinfo {title} {Whoi-plankton-a large scale fine grained visual recognition benchmark dataset for plankton classification},\ }\bibfield  {journal} {\bibinfo  {journal} {arXiv preprint arXiv:1510.00745}\ }\textbf {\bibinfo {volume} {abs/1510.00745}},\ \href {https://doi.org/10.48550/arXiv.1510.00745} {10.48550/arXiv.1510.00745} (\bibinfo {year} {2015}),\ \Eprint {https://arxiv.org/abs/1510.00745} {1510.00745} \BibitemShut {NoStop}%
\bibitem [{\citenamefont {Hsiang}\ \emph {et~al.}(2019)\citenamefont {Hsiang}, \citenamefont {Brombacher}, \citenamefont {Rillo}, \citenamefont {Mleneck-Vautravers}, \citenamefont {Conn}, \citenamefont {Lordsmith}, \citenamefont {Jentzen}, \citenamefont {Henehan}, \citenamefont {Metcalfe}, \citenamefont {Fenton}, \citenamefont {Wade}, \citenamefont {Fox}, \citenamefont {Meilland}, \citenamefont {Davis}, \citenamefont {Baranowski}, \citenamefont {Groeneveld}, \citenamefont {Edgar}, \citenamefont {Movellan}, \citenamefont {Aze}, \citenamefont {Dowsett}, \citenamefont {Miller}, \citenamefont {Rios},\ and\ \citenamefont {Hull}}]{hsiang_endless_2019}%
  \BibitemOpen
  \bibfield  {author} {\bibinfo {author} {\bibfnamefont {A.~Y.}\ \bibnamefont {Hsiang}}, \bibinfo {author} {\bibfnamefont {A.}~\bibnamefont {Brombacher}}, \bibinfo {author} {\bibfnamefont {M.~C.}\ \bibnamefont {Rillo}}, \bibinfo {author} {\bibfnamefont {M.~J.}\ \bibnamefont {Mleneck-Vautravers}}, \bibinfo {author} {\bibfnamefont {S.}~\bibnamefont {Conn}}, \bibinfo {author} {\bibfnamefont {S.}~\bibnamefont {Lordsmith}}, \bibinfo {author} {\bibfnamefont {A.}~\bibnamefont {Jentzen}}, \bibinfo {author} {\bibfnamefont {M.~J.}\ \bibnamefont {Henehan}}, \bibinfo {author} {\bibfnamefont {B.}~\bibnamefont {Metcalfe}}, \bibinfo {author} {\bibfnamefont {I.~S.}\ \bibnamefont {Fenton}}, \bibinfo {author} {\bibfnamefont {B.~S.}\ \bibnamefont {Wade}}, \bibinfo {author} {\bibfnamefont {L.}~\bibnamefont {Fox}}, \bibinfo {author} {\bibfnamefont {J.}~\bibnamefont {Meilland}}, \bibinfo {author} {\bibfnamefont {C.~V.}\ \bibnamefont {Davis}}, \bibinfo {author} {\bibfnamefont {U.}~\bibnamefont {Baranowski}}, \bibinfo {author} {\bibfnamefont {J.}~\bibnamefont {Groeneveld}}, \bibinfo {author} {\bibfnamefont {K.~M.}\ \bibnamefont {Edgar}}, \bibinfo {author} {\bibfnamefont {A.}~\bibnamefont {Movellan}}, \bibinfo {author} {\bibfnamefont {T.}~\bibnamefont {Aze}}, \bibinfo {author} {\bibfnamefont {H.~J.}\ \bibnamefont {Dowsett}}, \bibinfo {author} {\bibfnamefont {C.~G.}\ \bibnamefont {Miller}}, \bibinfo {author} {\bibfnamefont {N.}~\bibnamefont {Rios}},\ and\ \bibinfo {author} {\bibfnamefont {P.~M.}\ \bibnamefont {Hull}},\ }\bibfield  {title} {\bibinfo {title} {Endless {Forams}: {\textgreater}34,000 {Modern} {Planktonic} {Foraminiferal} {Images} for {Taxonomic} {Training} and {Automated} {Species} {Recognition} {Using} {Convolutional} {Neural} {Networks}},\ }\href {https://doi.org/https://doi.org/10.1029/2019PA003612} {\bibfield  {journal} {\bibinfo  {journal} {Paleoceanography and Paleoclimatology}\ }\textbf {\bibinfo {volume} {34}},\ \bibinfo {pages} {1157} (\bibinfo {year} {2019})}\BibitemShut {NoStop}%
\bibitem [{\citenamefont {Ohman}\ \emph {et~al.}(2019)\citenamefont {Ohman}, \citenamefont {Davis}, \citenamefont {Sherman}, \citenamefont {Grindley}, \citenamefont {Whitmore}, \citenamefont {Nickels},\ and\ \citenamefont {Ellen}}]{Ohman2019}%
  \BibitemOpen
  \bibfield  {author} {\bibinfo {author} {\bibfnamefont {M.~D.}\ \bibnamefont {Ohman}}, \bibinfo {author} {\bibfnamefont {R.~E.}\ \bibnamefont {Davis}}, \bibinfo {author} {\bibfnamefont {J.~T.}\ \bibnamefont {Sherman}}, \bibinfo {author} {\bibfnamefont {K.~R.}\ \bibnamefont {Grindley}}, \bibinfo {author} {\bibfnamefont {B.~M.}\ \bibnamefont {Whitmore}}, \bibinfo {author} {\bibfnamefont {C.~F.}\ \bibnamefont {Nickels}},\ and\ \bibinfo {author} {\bibfnamefont {J.~S.}\ \bibnamefont {Ellen}},\ }\bibfield  {title} {\bibinfo {title} {{Zooglider: An autonomous vehicle for optical and acoustic sensing of zooplankton}},\ }\href {https://doi.org/10.1002/lom3.10301} {\bibfield  {journal} {\bibinfo  {journal} {Limnology and Oceanography: Methods}\ }\textbf {\bibinfo {volume} {17}},\ \bibinfo {pages} {69} (\bibinfo {year} {2019})}\BibitemShut {NoStop}%
\bibitem [{\citenamefont {Gorsky}\ \emph {et~al.}(2010)\citenamefont {Gorsky}, \citenamefont {Ohman}, \citenamefont {Picheral}, \citenamefont {Gasparini}, \citenamefont {Stemmann}, \citenamefont {Romagnan}, \citenamefont {Cawood}, \citenamefont {Pesant}, \citenamefont {Garc{\'\i}a-Comas},\ and\ \citenamefont {Prejger}}]{gorsky2010digital}%
  \BibitemOpen
  \bibfield  {author} {\bibinfo {author} {\bibfnamefont {G.}~\bibnamefont {Gorsky}}, \bibinfo {author} {\bibfnamefont {M.~D.}\ \bibnamefont {Ohman}}, \bibinfo {author} {\bibfnamefont {M.}~\bibnamefont {Picheral}}, \bibinfo {author} {\bibfnamefont {S.}~\bibnamefont {Gasparini}}, \bibinfo {author} {\bibfnamefont {L.}~\bibnamefont {Stemmann}}, \bibinfo {author} {\bibfnamefont {J.-B.}\ \bibnamefont {Romagnan}}, \bibinfo {author} {\bibfnamefont {A.}~\bibnamefont {Cawood}}, \bibinfo {author} {\bibfnamefont {S.}~\bibnamefont {Pesant}}, \bibinfo {author} {\bibfnamefont {C.}~\bibnamefont {Garc{\'\i}a-Comas}},\ and\ \bibinfo {author} {\bibfnamefont {F.}~\bibnamefont {Prejger}},\ }\bibfield  {title} {\bibinfo {title} {Digital zooplankton image analysis using the zooscan integrated system},\ }\href@noop {} {\bibfield  {journal} {\bibinfo  {journal} {Journal of plankton research}\ }\textbf {\bibinfo {volume} {32}},\ \bibinfo {pages} {285} (\bibinfo {year} {2010})}\BibitemShut {NoStop}%
\bibitem [{\citenamefont {Picheral}\ \emph {et~al.}(2022)\citenamefont {Picheral}, \citenamefont {Catalano}, \citenamefont {Brousseau}, \citenamefont {Claustre}, \citenamefont {Coppola}, \citenamefont {Leymarie}, \citenamefont {Coindat}, \citenamefont {Dias}, \citenamefont {Fevre}, \citenamefont {Guidi} \emph {et~al.}}]{picheral2022underwater}%
  \BibitemOpen
  \bibfield  {author} {\bibinfo {author} {\bibfnamefont {M.}~\bibnamefont {Picheral}}, \bibinfo {author} {\bibfnamefont {C.}~\bibnamefont {Catalano}}, \bibinfo {author} {\bibfnamefont {D.}~\bibnamefont {Brousseau}}, \bibinfo {author} {\bibfnamefont {H.}~\bibnamefont {Claustre}}, \bibinfo {author} {\bibfnamefont {L.}~\bibnamefont {Coppola}}, \bibinfo {author} {\bibfnamefont {E.}~\bibnamefont {Leymarie}}, \bibinfo {author} {\bibfnamefont {J.}~\bibnamefont {Coindat}}, \bibinfo {author} {\bibfnamefont {F.}~\bibnamefont {Dias}}, \bibinfo {author} {\bibfnamefont {S.}~\bibnamefont {Fevre}}, \bibinfo {author} {\bibfnamefont {L.}~\bibnamefont {Guidi}}, \emph {et~al.},\ }\bibfield  {title} {\bibinfo {title} {The underwater vision profiler 6: an imaging sensor of particle size spectra and plankton, for autonomous and cabled platforms},\ }\href@noop {} {\bibfield  {journal} {\bibinfo  {journal} {Limnology and Oceanography: Methods}\ }\textbf {\bibinfo {volume} {20}},\ \bibinfo {pages} {115} (\bibinfo {year} {2022})}\BibitemShut {NoStop}%
\bibitem [{\citenamefont {Midtvedt}\ \emph {et~al.}(2022)\citenamefont {Midtvedt}, \citenamefont {Pineda}, \citenamefont {Skärberg}, \citenamefont {Olsén}, \citenamefont {Bachimanchi}, \citenamefont {Wesén}, \citenamefont {Esbjörner}, \citenamefont {Selander}, \citenamefont {Höök}, \citenamefont {Midtvedt},\ and\ \citenamefont {Volpe}}]{midtvedt_single-shot_2022}%
  \BibitemOpen
  \bibfield  {author} {\bibinfo {author} {\bibfnamefont {B.}~\bibnamefont {Midtvedt}}, \bibinfo {author} {\bibfnamefont {J.}~\bibnamefont {Pineda}}, \bibinfo {author} {\bibfnamefont {F.}~\bibnamefont {Skärberg}}, \bibinfo {author} {\bibfnamefont {E.}~\bibnamefont {Olsén}}, \bibinfo {author} {\bibfnamefont {H.}~\bibnamefont {Bachimanchi}}, \bibinfo {author} {\bibfnamefont {E.}~\bibnamefont {Wesén}}, \bibinfo {author} {\bibfnamefont {E.~K.}\ \bibnamefont {Esbjörner}}, \bibinfo {author} {\bibfnamefont {E.}~\bibnamefont {Selander}}, \bibinfo {author} {\bibfnamefont {F.}~\bibnamefont {Höök}}, \bibinfo {author} {\bibfnamefont {D.}~\bibnamefont {Midtvedt}},\ and\ \bibinfo {author} {\bibfnamefont {G.}~\bibnamefont {Volpe}},\ }\bibfield  {title} {{\selectlanguage {english}\bibinfo {title} {Single-shot self-supervised object detection in microscopy}},\ }\href {https://doi.org/10.1038/s41467-022-35004-y} {\bibfield  {journal} {\bibinfo  {journal} {Nature Communications}\ }\textbf {\bibinfo {volume} {13}},\ \bibinfo {pages} {7492} (\bibinfo {year} {2022})},\ \bibinfo {note} {number: 1 Publisher: Nature Publishing Group}\BibitemShut {NoStop}%
\bibitem [{\citenamefont {Girshick}\ \emph {et~al.}(2013)\citenamefont {Girshick}, \citenamefont {Donahue}, \citenamefont {Darrell},\ and\ \citenamefont {Malik}}]{Girshick2013}%
  \BibitemOpen
  \bibfield  {author} {\bibinfo {author} {\bibfnamefont {R.}~\bibnamefont {Girshick}}, \bibinfo {author} {\bibfnamefont {J.}~\bibnamefont {Donahue}}, \bibinfo {author} {\bibfnamefont {T.}~\bibnamefont {Darrell}},\ and\ \bibinfo {author} {\bibfnamefont {J.}~\bibnamefont {Malik}},\ }\bibfield  {title} {\bibinfo {title} {{Rich feature hierarchies for accurate object detection and semantic segmentation}},\ }\bibfield  {journal} {\bibinfo  {journal} {arXiv}\ }\href {https://doi.org/10.48550/ARXIV.1311.2524} {10.48550/ARXIV.1311.2524} (\bibinfo {year} {2013})\BibitemShut {NoStop}%
\bibitem [{\citenamefont {Ren}\ \emph {et~al.}(2015)\citenamefont {Ren}, \citenamefont {He}, \citenamefont {Girshick},\ and\ \citenamefont {Sun}}]{Ren2015}%
  \BibitemOpen
  \bibfield  {author} {\bibinfo {author} {\bibfnamefont {S.}~\bibnamefont {Ren}}, \bibinfo {author} {\bibfnamefont {K.}~\bibnamefont {He}}, \bibinfo {author} {\bibfnamefont {R.}~\bibnamefont {Girshick}},\ and\ \bibinfo {author} {\bibfnamefont {J.}~\bibnamefont {Sun}},\ }\bibfield  {title} {\bibinfo {title} {Faster r-cnn: Towards real-time object detection with region proposal networks},\ }\href@noop {} {\bibfield  {journal} {\bibinfo  {journal} {Advances in neural information processing systems}\ }\textbf {\bibinfo {volume} {28}} (\bibinfo {year} {2015})}\BibitemShut {NoStop}%
\bibitem [{\citenamefont {Redmon}\ \emph {et~al.}(2016)\citenamefont {Redmon}, \citenamefont {Divvala}, \citenamefont {Girshick},\ and\ \citenamefont {Farhadi}}]{Redmon2016}%
  \BibitemOpen
  \bibfield  {author} {\bibinfo {author} {\bibfnamefont {J.}~\bibnamefont {Redmon}}, \bibinfo {author} {\bibfnamefont {S.}~\bibnamefont {Divvala}}, \bibinfo {author} {\bibfnamefont {R.}~\bibnamefont {Girshick}},\ and\ \bibinfo {author} {\bibfnamefont {A.}~\bibnamefont {Farhadi}},\ }\bibfield  {title} {\bibinfo {title} {You only look once: Unified, real-time object detection},\ }in\ \href@noop {} {\emph {\bibinfo {booktitle} {Proceedings of the IEEE conference on computer vision and pattern recognition}}}\ (\bibinfo {year} {2016})\ pp.\ \bibinfo {pages} {779--788}\BibitemShut {NoStop}%
\bibitem [{\citenamefont {Pedraza}\ \emph {et~al.}(2018)\citenamefont {Pedraza}, \citenamefont {Bueno}, \citenamefont {Deniz}, \citenamefont {Ruiz-Santaquiteria}, \citenamefont {Sanchez}, \citenamefont {Blanco}, \citenamefont {Borrego-Ramos}, \citenamefont {Olenici},\ and\ \citenamefont {Cristobal}}]{Pedraza2018}%
  \BibitemOpen
  \bibfield  {author} {\bibinfo {author} {\bibfnamefont {A.}~\bibnamefont {Pedraza}}, \bibinfo {author} {\bibfnamefont {G.}~\bibnamefont {Bueno}}, \bibinfo {author} {\bibfnamefont {O.}~\bibnamefont {Deniz}}, \bibinfo {author} {\bibfnamefont {J.}~\bibnamefont {Ruiz-Santaquiteria}}, \bibinfo {author} {\bibfnamefont {C.}~\bibnamefont {Sanchez}}, \bibinfo {author} {\bibfnamefont {S.}~\bibnamefont {Blanco}}, \bibinfo {author} {\bibfnamefont {M.}~\bibnamefont {Borrego-Ramos}}, \bibinfo {author} {\bibfnamefont {A.}~\bibnamefont {Olenici}},\ and\ \bibinfo {author} {\bibfnamefont {G.}~\bibnamefont {Cristobal}},\ }\bibfield  {title} {\bibinfo {title} {{Lights and pitfalls of convolutional neural networks for diatom identification}},\ }in\ \href {https://doi.org/10.1117/12.2309488} {\emph {\bibinfo {booktitle} {Proc.SPIE}}},\ Vol.\ \bibinfo {volume} {10679}\ (\bibinfo {year} {2018})\BibitemShut {NoStop}%
\bibitem [{\citenamefont {Kakehi}\ \emph {et~al.}(2021)\citenamefont {Kakehi}, \citenamefont {Sekiuchi}, \citenamefont {Ito}, \citenamefont {Ueno}, \citenamefont {Takeuchi}, \citenamefont {Suzuki},\ and\ \citenamefont {Togawa}}]{kakehi_identification_2021}%
  \BibitemOpen
  \bibfield  {author} {\bibinfo {author} {\bibfnamefont {S.}~\bibnamefont {Kakehi}}, \bibinfo {author} {\bibfnamefont {T.}~\bibnamefont {Sekiuchi}}, \bibinfo {author} {\bibfnamefont {H.}~\bibnamefont {Ito}}, \bibinfo {author} {\bibfnamefont {S.}~\bibnamefont {Ueno}}, \bibinfo {author} {\bibfnamefont {Y.}~\bibnamefont {Takeuchi}}, \bibinfo {author} {\bibfnamefont {K.}~\bibnamefont {Suzuki}},\ and\ \bibinfo {author} {\bibfnamefont {M.}~\bibnamefont {Togawa}},\ }\bibfield  {title} {\bibinfo {title} {Identification and counting of {Pacific} oyster {Crassostrea} gigas larvae by object detection using deep learning},\ }\href {https://doi.org/10.1016/j.aquaeng.2021.102197} {\bibfield  {journal} {\bibinfo  {journal} {Aquacultural Engineering}\ }\textbf {\bibinfo {volume} {95}},\ \bibinfo {pages} {102197} (\bibinfo {year} {2021})}\BibitemShut {NoStop}%
\bibitem [{\citenamefont {Wu}\ \emph {et~al.}(2019)\citenamefont {Wu}, \citenamefont {Kirillov}, \citenamefont {Massa}, \citenamefont {Lo},\ and\ \citenamefont {Girshick}}]{wu2019detectron2}%
  \BibitemOpen
  \bibfield  {author} {\bibinfo {author} {\bibfnamefont {Y.}~\bibnamefont {Wu}}, \bibinfo {author} {\bibfnamefont {A.}~\bibnamefont {Kirillov}}, \bibinfo {author} {\bibfnamefont {F.}~\bibnamefont {Massa}}, \bibinfo {author} {\bibfnamefont {W.-Y.}\ \bibnamefont {Lo}},\ and\ \bibinfo {author} {\bibfnamefont {R.}~\bibnamefont {Girshick}},\ }\href@noop {} {\bibinfo {title} {Detectron2}},\ \bibinfo {howpublished} {\url{https://github.com/facebookresearch/detectron2}} (\bibinfo {year} {2019})\BibitemShut {NoStop}%
\bibitem [{\citenamefont {Ruiz-Santaquiteria}\ \emph {et~al.}(2020)\citenamefont {Ruiz-Santaquiteria}, \citenamefont {Bueno}, \citenamefont {Deniz}, \citenamefont {Vallez},\ and\ \citenamefont {Cristobal}}]{RUIZSANTAQUITERIA2020103271}%
  \BibitemOpen
  \bibfield  {author} {\bibinfo {author} {\bibfnamefont {J.}~\bibnamefont {Ruiz-Santaquiteria}}, \bibinfo {author} {\bibfnamefont {G.}~\bibnamefont {Bueno}}, \bibinfo {author} {\bibfnamefont {O.}~\bibnamefont {Deniz}}, \bibinfo {author} {\bibfnamefont {N.}~\bibnamefont {Vallez}},\ and\ \bibinfo {author} {\bibfnamefont {G.}~\bibnamefont {Cristobal}},\ }\bibfield  {title} {\bibinfo {title} {Semantic versus instance segmentation in microscopic algae detection},\ }\href {https://doi.org/https://doi.org/10.1016/j.engappai.2019.103271} {\bibfield  {journal} {\bibinfo  {journal} {Engineering Applications of Artificial Intelligence}\ }\textbf {\bibinfo {volume} {87}},\ \bibinfo {pages} {103271} (\bibinfo {year} {2020})}\BibitemShut {NoStop}%
\bibitem [{\citenamefont {He}\ \emph {et~al.}(2017)\citenamefont {He}, \citenamefont {Gkioxari}, \citenamefont {Doll{\'a}r},\ and\ \citenamefont {Girshick}}]{He2017}%
  \BibitemOpen
  \bibfield  {author} {\bibinfo {author} {\bibfnamefont {K.}~\bibnamefont {He}}, \bibinfo {author} {\bibfnamefont {G.}~\bibnamefont {Gkioxari}}, \bibinfo {author} {\bibfnamefont {P.}~\bibnamefont {Doll{\'a}r}},\ and\ \bibinfo {author} {\bibfnamefont {R.}~\bibnamefont {Girshick}},\ }\bibfield  {title} {\bibinfo {title} {Mask r-cnn},\ }in\ \href@noop {} {\emph {\bibinfo {booktitle} {Proceedings of the IEEE international conference on computer vision}}}\ (\bibinfo {year} {2017})\ pp.\ \bibinfo {pages} {2961--2969}\BibitemShut {NoStop}%
\bibitem [{\citenamefont {Badrinarayanan}\ \emph {et~al.}(2017)\citenamefont {Badrinarayanan}, \citenamefont {Kendall},\ and\ \citenamefont {Cipolla}}]{SegNet}%
  \BibitemOpen
  \bibfield  {author} {\bibinfo {author} {\bibfnamefont {V.}~\bibnamefont {Badrinarayanan}}, \bibinfo {author} {\bibfnamefont {A.}~\bibnamefont {Kendall}},\ and\ \bibinfo {author} {\bibfnamefont {R.}~\bibnamefont {Cipolla}},\ }\href@noop {} {\bibinfo {title} {Segnet: A deep convolutional encoder-decoder architecture for image segmentation}} (\bibinfo {year} {2017})\BibitemShut {NoStop}%
\bibitem [{\citenamefont {Shi}\ \emph {et~al.}(2019)\citenamefont {Shi}, \citenamefont {Wang}, \citenamefont {Cao}, \citenamefont {Ren}, \citenamefont {Han},\ and\ \citenamefont {Ma}}]{shi2019study}%
  \BibitemOpen
  \bibfield  {author} {\bibinfo {author} {\bibfnamefont {Z.}~\bibnamefont {Shi}}, \bibinfo {author} {\bibfnamefont {K.}~\bibnamefont {Wang}}, \bibinfo {author} {\bibfnamefont {L.}~\bibnamefont {Cao}}, \bibinfo {author} {\bibfnamefont {Y.}~\bibnamefont {Ren}}, \bibinfo {author} {\bibfnamefont {Y.}~\bibnamefont {Han}},\ and\ \bibinfo {author} {\bibfnamefont {S.}~\bibnamefont {Ma}},\ }\bibfield  {title} {\bibinfo {title} {Study on holographic image recognition technology of zooplankton},\ }\href@noop {} {\bibfield  {journal} {\bibinfo  {journal} {DEStech Trans. Comput. Sci. Eng}\ ,\ \bibinfo {pages} {580}} (\bibinfo {year} {2019})}\BibitemShut {NoStop}%
\bibitem [{\citenamefont {Li}\ \emph {et~al.}(2021)\citenamefont {Li}, \citenamefont {Guo}, \citenamefont {Guo}, \citenamefont {Hu},\ and\ \citenamefont {Tian}}]{li_plankton_2021}%
  \BibitemOpen
  \bibfield  {author} {\bibinfo {author} {\bibfnamefont {Y.}~\bibnamefont {Li}}, \bibinfo {author} {\bibfnamefont {J.}~\bibnamefont {Guo}}, \bibinfo {author} {\bibfnamefont {X.}~\bibnamefont {Guo}}, \bibinfo {author} {\bibfnamefont {Z.}~\bibnamefont {Hu}},\ and\ \bibinfo {author} {\bibfnamefont {Y.}~\bibnamefont {Tian}},\ }\bibfield  {title} {\bibinfo {title} {Plankton {detection} with {adversarial} {learning} and a {densely} {connected} {deep} {learning} {model} for {class} {imbalanced} {distribution}},\ }\bibfield  {journal} {\bibinfo  {journal} {Journal of Marine Science and Engineering}\ }\textbf {\bibinfo {volume} {9}},\ \href {https://doi.org/10.3390/jmse9060636} {10.3390/jmse9060636} (\bibinfo {year} {2021})\BibitemShut {NoStop}%
\bibitem [{\citenamefont {Huang}\ \emph {et~al.}(2016)\citenamefont {Huang}, \citenamefont {Liu}, \citenamefont {van~der Maaten},\ and\ \citenamefont {Weinberger}}]{huang2016}%
  \BibitemOpen
  \bibfield  {author} {\bibinfo {author} {\bibfnamefont {G.}~\bibnamefont {Huang}}, \bibinfo {author} {\bibfnamefont {Z.}~\bibnamefont {Liu}}, \bibinfo {author} {\bibfnamefont {L.}~\bibnamefont {van~der Maaten}},\ and\ \bibinfo {author} {\bibfnamefont {K.~Q.}\ \bibnamefont {Weinberger}},\ }\href {https://doi.org/10.48550/ARXIV.1608.06993} {\bibinfo {title} {{Densely Connected Convolutional Networks}}} (\bibinfo {year} {2016})\BibitemShut {NoStop}%
\bibitem [{\citenamefont {Helgadottir}\ \emph {et~al.}(2019)\citenamefont {Helgadottir}, \citenamefont {Argun},\ and\ \citenamefont {Volpe}}]{helgadottir2019digital}%
  \BibitemOpen
  \bibfield  {author} {\bibinfo {author} {\bibfnamefont {S.}~\bibnamefont {Helgadottir}}, \bibinfo {author} {\bibfnamefont {A.}~\bibnamefont {Argun}},\ and\ \bibinfo {author} {\bibfnamefont {G.}~\bibnamefont {Volpe}},\ }\bibfield  {title} {\bibinfo {title} {Digital video microscopy enhanced by deep learning},\ }\href {https://doi.org/10.1364/OPTICA.6.000506} {\bibfield  {journal} {\bibinfo  {journal} {Optica}\ }\textbf {\bibinfo {volume} {6}},\ \bibinfo {pages} {506} (\bibinfo {year} {2019})}\BibitemShut {NoStop}%
\bibitem [{\citenamefont {Midtvedt}\ \emph {et~al.}(2021)\citenamefont {Midtvedt}, \citenamefont {Helgadottir}, \citenamefont {Argun}, \citenamefont {Pineda}, \citenamefont {Midtvedt},\ and\ \citenamefont {Volpe}}]{midtvedt2021}%
  \BibitemOpen
  \bibfield  {author} {\bibinfo {author} {\bibfnamefont {B.}~\bibnamefont {Midtvedt}}, \bibinfo {author} {\bibfnamefont {S.}~\bibnamefont {Helgadottir}}, \bibinfo {author} {\bibfnamefont {A.}~\bibnamefont {Argun}}, \bibinfo {author} {\bibfnamefont {J.}~\bibnamefont {Pineda}}, \bibinfo {author} {\bibfnamefont {D.}~\bibnamefont {Midtvedt}},\ and\ \bibinfo {author} {\bibfnamefont {G.}~\bibnamefont {Volpe}},\ }\bibfield  {title} {\bibinfo {title} {{Quantitative digital microscopy with deep learning}},\ }\href {https://doi.org/10.1063/5.0034891} {\bibfield  {journal} {\bibinfo  {journal} {Applied Physics Reviews}\ }\textbf {\bibinfo {volume} {8}},\ \bibinfo {pages} {011310} (\bibinfo {year} {2021})},\ \Eprint {https://arxiv.org/abs/https://pubs.aip.org/aip/apr/article-pdf/doi/10.1063/5.0034891/14577703/011310\_1\_online.pdf} {https://pubs.aip.org/aip/apr/article-pdf/doi/10.1063/5.0034891/14577703/011310\_1\_online.pdf} \BibitemShut {NoStop}%
\bibitem [{\citenamefont {Kuang}(2015)}]{kuang_deep_2015}%
  \BibitemOpen
  \bibfield  {author} {\bibinfo {author} {\bibfnamefont {Y.}~\bibnamefont {Kuang}},\ }\href@noop {} {\emph {\bibinfo {title} {Deep neural network for deep sea plankton classification}}},\ \bibinfo {type} {Tech. Rep.}\ (\bibinfo  {institution} {Technical Report 2015. Available online: https://pdfs. semanticscholar. org~…},\ \bibinfo {year} {2015})\BibitemShut {NoStop}%
\bibitem [{\citenamefont {Dai}\ \emph {et~al.}(2016)\citenamefont {Dai}, \citenamefont {Wang}, \citenamefont {Zheng}, \citenamefont {Ji},\ and\ \citenamefont {Qiao}}]{dai_zooplanktonet_2016}%
  \BibitemOpen
  \bibfield  {author} {\bibinfo {author} {\bibfnamefont {J.}~\bibnamefont {Dai}}, \bibinfo {author} {\bibfnamefont {R.}~\bibnamefont {Wang}}, \bibinfo {author} {\bibfnamefont {H.}~\bibnamefont {Zheng}}, \bibinfo {author} {\bibfnamefont {G.}~\bibnamefont {Ji}},\ and\ \bibinfo {author} {\bibfnamefont {X.}~\bibnamefont {Qiao}},\ }\bibfield  {title} {\bibinfo {title} {{ZooplanktoNet}: {Deep} convolutional network for zooplankton classification},\ }in\ \href {https://doi.org/10.1109/OCEANSAP.2016.7485680} {\emph {\bibinfo {booktitle} {{OCEANS} 2016 - {Shanghai}}}}\ (\bibinfo {organization} {IEEE},\ \bibinfo {year} {2016})\ pp.\ \bibinfo {pages} {1--6}\BibitemShut {NoStop}%
\bibitem [{\citenamefont {Correa}\ \emph {et~al.}(2017)\citenamefont {Correa}, \citenamefont {Drews}, \citenamefont {Botelho}, \citenamefont {Souza},\ and\ \citenamefont {Tavano}}]{correa_deep_2017}%
  \BibitemOpen
  \bibfield  {author} {\bibinfo {author} {\bibfnamefont {I.}~\bibnamefont {Correa}}, \bibinfo {author} {\bibfnamefont {P.}~\bibnamefont {Drews}}, \bibinfo {author} {\bibfnamefont {S.}~\bibnamefont {Botelho}}, \bibinfo {author} {\bibfnamefont {M.~S.~d.}\ \bibnamefont {Souza}},\ and\ \bibinfo {author} {\bibfnamefont {V.~M.}\ \bibnamefont {Tavano}},\ }\bibfield  {title} {\bibinfo {title} {Deep {Learning} for {Microalgae} {Classification}},\ }in\ \href {https://doi.org/10.1109/ICMLA.2017.0-183} {\emph {\bibinfo {booktitle} {2017 16th {IEEE} {International} {Conference} on {Machine} {Learning} and {Applications} ({ICMLA})}}}\ (\bibinfo {organization} {IEEE},\ \bibinfo {year} {2017})\ pp.\ \bibinfo {pages} {20--25}\BibitemShut {NoStop}%
\bibitem [{\citenamefont {Luo}\ \emph {et~al.}(2018)\citenamefont {Luo}, \citenamefont {Irisson}, \citenamefont {Graham}, \citenamefont {Guigand}, \citenamefont {Sarafraz}, \citenamefont {Mader},\ and\ \citenamefont {Cowen}}]{luo_automated_2018}%
  \BibitemOpen
  \bibfield  {author} {\bibinfo {author} {\bibfnamefont {J.~Y.}\ \bibnamefont {Luo}}, \bibinfo {author} {\bibfnamefont {J.-O.}\ \bibnamefont {Irisson}}, \bibinfo {author} {\bibfnamefont {B.}~\bibnamefont {Graham}}, \bibinfo {author} {\bibfnamefont {C.}~\bibnamefont {Guigand}}, \bibinfo {author} {\bibfnamefont {A.}~\bibnamefont {Sarafraz}}, \bibinfo {author} {\bibfnamefont {C.}~\bibnamefont {Mader}},\ and\ \bibinfo {author} {\bibfnamefont {R.~K.}\ \bibnamefont {Cowen}},\ }\bibfield  {title} {\bibinfo {title} {Automated plankton image analysis using convolutional neural networks},\ }\href {https://doi.org/10.1002/lom3.10285} {\bibfield  {journal} {\bibinfo  {journal} {Limnology and Oceanography: Methods}\ }\textbf {\bibinfo {volume} {16}},\ \bibinfo {pages} {814} (\bibinfo {year} {2018})}\BibitemShut {NoStop}%
\bibitem [{\citenamefont {Cowen}\ and\ \citenamefont {Guigand}(2008)}]{Cowen2008}%
  \BibitemOpen
  \bibfield  {author} {\bibinfo {author} {\bibfnamefont {R.~K.}\ \bibnamefont {Cowen}}\ and\ \bibinfo {author} {\bibfnamefont {C.~M.}\ \bibnamefont {Guigand}},\ }\bibfield  {title} {\bibinfo {title} {{In situ ichthyoplankton imaging system (ISIIS): system design and preliminary results}},\ }\href {https://doi.org/10.4319/lom.2008.6.126} {\bibfield  {journal} {\bibinfo  {journal} {Limnology and Oceanography: Methods}\ }\textbf {\bibinfo {volume} {6}},\ \bibinfo {pages} {126} (\bibinfo {year} {2008})}\BibitemShut {NoStop}%
\bibitem [{\citenamefont {Zhu}\ \emph {et~al.}(2020)\citenamefont {Zhu}, \citenamefont {Park}, \citenamefont {Isola},\ and\ \citenamefont {Efros}}]{Zhu2020}%
  \BibitemOpen
  \bibfield  {author} {\bibinfo {author} {\bibfnamefont {J.-Y.}\ \bibnamefont {Zhu}}, \bibinfo {author} {\bibfnamefont {T.}~\bibnamefont {Park}}, \bibinfo {author} {\bibfnamefont {P.}~\bibnamefont {Isola}},\ and\ \bibinfo {author} {\bibfnamefont {A.~A.}\ \bibnamefont {Efros}},\ }\href@noop {} {\bibinfo {title} {Unpaired image-to-image translation using cycle-consistent adversarial networks}} (\bibinfo {year} {2020}),\ \Eprint {https://arxiv.org/abs/1703.10593} {arXiv:1703.10593 [cs.CV]} \BibitemShut {NoStop}%
\bibitem [{\citenamefont {Lee}\ \emph {et~al.}(2016)\citenamefont {Lee}, \citenamefont {Park},\ and\ \citenamefont {Kim}}]{lee2016plankton}%
  \BibitemOpen
  \bibfield  {author} {\bibinfo {author} {\bibfnamefont {H.}~\bibnamefont {Lee}}, \bibinfo {author} {\bibfnamefont {M.}~\bibnamefont {Park}},\ and\ \bibinfo {author} {\bibfnamefont {J.}~\bibnamefont {Kim}},\ }\bibfield  {title} {\bibinfo {title} {Plankton classification on imbalanced large scale database via convolutional neural networks with transfer learning},\ }in\ \href@noop {} {\emph {\bibinfo {booktitle} {2016 IEEE international conference on image processing (ICIP)}}}\ (\bibinfo {organization} {IEEE},\ \bibinfo {year} {2016})\ pp.\ \bibinfo {pages} {3713--3717}\BibitemShut {NoStop}%
\bibitem [{\citenamefont {Deng}\ \emph {et~al.}(2009)\citenamefont {Deng}, \citenamefont {Dong}, \citenamefont {Socher}, \citenamefont {Li}, \citenamefont {{Kai Li}},\ and\ \citenamefont {{Li Fei-Fei}}}]{Deng2009}%
  \BibitemOpen
  \bibfield  {author} {\bibinfo {author} {\bibfnamefont {J.}~\bibnamefont {Deng}}, \bibinfo {author} {\bibfnamefont {W.}~\bibnamefont {Dong}}, \bibinfo {author} {\bibfnamefont {R.}~\bibnamefont {Socher}}, \bibinfo {author} {\bibfnamefont {L.-J.}\ \bibnamefont {Li}}, \bibinfo {author} {\bibnamefont {{Kai Li}}},\ and\ \bibinfo {author} {\bibnamefont {{Li Fei-Fei}}},\ }\bibfield  {title} {\bibinfo {title} {{ImageNet: A large-scale hierarchical image database}},\ }in\ \href {https://doi.org/10.1109/cvpr.2009.5206848} {\emph {\bibinfo {booktitle} {2009 IEEE Conference on Computer Vision and Pattern Recognition}}}\ (\bibinfo {year} {2009})\ pp.\ \bibinfo {pages} {248--255}\BibitemShut {NoStop}%
\bibitem [{\citenamefont {Orenstein}\ and\ \citenamefont {Beijbom}(2017)}]{orenstein2017transfer}%
  \BibitemOpen
  \bibfield  {author} {\bibinfo {author} {\bibfnamefont {E.~C.}\ \bibnamefont {Orenstein}}\ and\ \bibinfo {author} {\bibfnamefont {O.}~\bibnamefont {Beijbom}},\ }\bibfield  {title} {\bibinfo {title} {Transfer learning and deep feature extraction for planktonic image data sets},\ }in\ \href@noop {} {\emph {\bibinfo {booktitle} {2017 IEEE Winter Conference on Applications of Computer Vision (WACV)}}}\ (\bibinfo {organization} {IEEE},\ \bibinfo {year} {2017})\ pp.\ \bibinfo {pages} {1082--1088}\BibitemShut {NoStop}%
\bibitem [{\citenamefont {MacNeil}\ \emph {et~al.}(2021)\citenamefont {MacNeil}, \citenamefont {Missan}, \citenamefont {Luo}, \citenamefont {Trappenberg},\ and\ \citenamefont {LaRoche}}]{macneil_plankton_2021}%
  \BibitemOpen
  \bibfield  {author} {\bibinfo {author} {\bibfnamefont {L.}~\bibnamefont {MacNeil}}, \bibinfo {author} {\bibfnamefont {S.}~\bibnamefont {Missan}}, \bibinfo {author} {\bibfnamefont {J.}~\bibnamefont {Luo}}, \bibinfo {author} {\bibfnamefont {T.}~\bibnamefont {Trappenberg}},\ and\ \bibinfo {author} {\bibfnamefont {J.}~\bibnamefont {LaRoche}},\ }\bibfield  {title} {\bibinfo {title} {Plankton classification with high-throughput submersible holographic microscopy and transfer learning},\ }\href {https://doi.org/10.1186/s12862-021-01839-0} {\bibfield  {journal} {\bibinfo  {journal} {BMC Ecology and Evolution}\ }\textbf {\bibinfo {volume} {21}},\ \bibinfo {pages} {123} (\bibinfo {year} {2021})}\BibitemShut {NoStop}%
\bibitem [{\citenamefont {Ellen}\ \emph {et~al.}(2019)\citenamefont {Ellen}, \citenamefont {Graff},\ and\ \citenamefont {Ohman}}]{ellen_improving_2019}%
  \BibitemOpen
  \bibfield  {author} {\bibinfo {author} {\bibfnamefont {J.~S.}\ \bibnamefont {Ellen}}, \bibinfo {author} {\bibfnamefont {C.~A.}\ \bibnamefont {Graff}},\ and\ \bibinfo {author} {\bibfnamefont {M.~D.}\ \bibnamefont {Ohman}},\ }\bibfield  {title} {\bibinfo {title} {Improving plankton image classification using context metadata},\ }\href {https://doi.org/10.1002/lom3.10324} {\bibfield  {journal} {\bibinfo  {journal} {Limnology and Oceanography: Methods}\ }\textbf {\bibinfo {volume} {17}},\ \bibinfo {pages} {439} (\bibinfo {year} {2019})}\BibitemShut {NoStop}%
\bibitem [{\citenamefont {Pinti}\ \emph {et~al.}(2023)\citenamefont {Pinti}, \citenamefont {J{\'o}nasd{\'o}ttir}, \citenamefont {Record},\ and\ \citenamefont {Visser}}]{pinti2023global}%
  \BibitemOpen
  \bibfield  {author} {\bibinfo {author} {\bibfnamefont {J.}~\bibnamefont {Pinti}}, \bibinfo {author} {\bibfnamefont {S.~H.}\ \bibnamefont {J{\'o}nasd{\'o}ttir}}, \bibinfo {author} {\bibfnamefont {N.~R.}\ \bibnamefont {Record}},\ and\ \bibinfo {author} {\bibfnamefont {A.~W.}\ \bibnamefont {Visser}},\ }\bibfield  {title} {\bibinfo {title} {The global contribution of seasonally migrating copepods to the biological carbon pump},\ }\href {https://doi.org/10.1002/lno.12335} {\bibfield  {journal} {\bibinfo  {journal} {Limnology and Oceanography}\ }\textbf {\bibinfo {volume} {68}},\ \bibinfo {pages} {1147} (\bibinfo {year} {2023})}\BibitemShut {NoStop}%
\bibitem [{\citenamefont {Bandara}\ \emph {et~al.}(2021)\citenamefont {Bandara}, \citenamefont {Varpe}, \citenamefont {Wijewardene}, \citenamefont {Tverberg},\ and\ \citenamefont {Eiane}}]{bandara2021two}%
  \BibitemOpen
  \bibfield  {author} {\bibinfo {author} {\bibfnamefont {K.}~\bibnamefont {Bandara}}, \bibinfo {author} {\bibfnamefont {{\O}.}~\bibnamefont {Varpe}}, \bibinfo {author} {\bibfnamefont {L.}~\bibnamefont {Wijewardene}}, \bibinfo {author} {\bibfnamefont {V.}~\bibnamefont {Tverberg}},\ and\ \bibinfo {author} {\bibfnamefont {K.}~\bibnamefont {Eiane}},\ }\bibfield  {title} {\bibinfo {title} {Two hundred years of zooplankton vertical migration research},\ }\href {https://doi.org/10.1111/brv.12715} {\bibfield  {journal} {\bibinfo  {journal} {Biological Reviews}\ }\textbf {\bibinfo {volume} {96}},\ \bibinfo {pages} {1547} (\bibinfo {year} {2021})}\BibitemShut {NoStop}%
\bibitem [{\citenamefont {Krishnamurthy}\ \emph {et~al.}(2020)\citenamefont {Krishnamurthy}, \citenamefont {Li}, \citenamefont {Benoit~du Rey}, \citenamefont {Cambournac}, \citenamefont {Larson}, \citenamefont {Li},\ and\ \citenamefont {Prakash}}]{krishnamurthy_scale-free_2020}%
  \BibitemOpen
  \bibfield  {author} {\bibinfo {author} {\bibfnamefont {D.}~\bibnamefont {Krishnamurthy}}, \bibinfo {author} {\bibfnamefont {H.}~\bibnamefont {Li}}, \bibinfo {author} {\bibfnamefont {F.}~\bibnamefont {Benoit~du Rey}}, \bibinfo {author} {\bibfnamefont {P.}~\bibnamefont {Cambournac}}, \bibinfo {author} {\bibfnamefont {A.~G.}\ \bibnamefont {Larson}}, \bibinfo {author} {\bibfnamefont {E.}~\bibnamefont {Li}},\ and\ \bibinfo {author} {\bibfnamefont {M.}~\bibnamefont {Prakash}},\ }\bibfield  {title} {{\selectlanguage {english}\bibinfo {title} {Scale-free vertical tracking microscopy}},\ }\href {https://doi.org/10.1038/s41592-020-0924-7} {\bibfield  {journal} {\bibinfo  {journal} {Nature Methods}\ }\textbf {\bibinfo {volume} {17}},\ \bibinfo {pages} {1040} (\bibinfo {year} {2020})},\ \bibinfo {note} {number: 10 Publisher: Nature Publishing Group}\BibitemShut {NoStop}%
\bibitem [{\citenamefont {Visser}\ and\ \citenamefont {Kiørboe}(2006)}]{visser_plankton_2006}%
  \BibitemOpen
  \bibfield  {author} {\bibinfo {author} {\bibfnamefont {A.~W.}\ \bibnamefont {Visser}}\ and\ \bibinfo {author} {\bibfnamefont {T.}~\bibnamefont {Kiørboe}},\ }\bibfield  {title} {{\selectlanguage {english}\bibinfo {title} {Plankton motility patterns and encounter rates}},\ }\href {https://doi.org/10.1007/s00442-006-0385-4} {\bibfield  {journal} {\bibinfo  {journal} {Oecologia}\ }\textbf {\bibinfo {volume} {148}},\ \bibinfo {pages} {538} (\bibinfo {year} {2006})}\BibitemShut {NoStop}%
\bibitem [{\citenamefont {Mu{\~n}oz-Gil}\ \emph {et~al.}(2021)\citenamefont {Mu{\~n}oz-Gil}, \citenamefont {Volpe}, \citenamefont {Garcia-March}, \citenamefont {Aghion}, \citenamefont {Argun}, \citenamefont {Hong}, \citenamefont {Bland}, \citenamefont {Bo}, \citenamefont {Conejero}, \citenamefont {Firbas} \emph {et~al.}}]{munoz2021objective}%
  \BibitemOpen
  \bibfield  {author} {\bibinfo {author} {\bibfnamefont {G.}~\bibnamefont {Mu{\~n}oz-Gil}}, \bibinfo {author} {\bibfnamefont {G.}~\bibnamefont {Volpe}}, \bibinfo {author} {\bibfnamefont {M.~A.}\ \bibnamefont {Garcia-March}}, \bibinfo {author} {\bibfnamefont {E.}~\bibnamefont {Aghion}}, \bibinfo {author} {\bibfnamefont {A.}~\bibnamefont {Argun}}, \bibinfo {author} {\bibfnamefont {C.~B.}\ \bibnamefont {Hong}}, \bibinfo {author} {\bibfnamefont {T.}~\bibnamefont {Bland}}, \bibinfo {author} {\bibfnamefont {S.}~\bibnamefont {Bo}}, \bibinfo {author} {\bibfnamefont {J.~A.}\ \bibnamefont {Conejero}}, \bibinfo {author} {\bibfnamefont {N.}~\bibnamefont {Firbas}}, \emph {et~al.},\ }\bibfield  {title} {\bibinfo {title} {Objective comparison of methods to decode anomalous diffusion},\ }\href@noop {} {\bibfield  {journal} {\bibinfo  {journal} {Nature communications}\ }\textbf {\bibinfo {volume} {12}},\ \bibinfo {pages} {6253} (\bibinfo {year} {2021})}\BibitemShut {NoStop}%
\bibitem [{\citenamefont {Bartumeus}\ \emph {et~al.}(2003)\citenamefont {Bartumeus}, \citenamefont {Peters}, \citenamefont {Pueyo}, \citenamefont {Marrasé},\ and\ \citenamefont {Catalan}}]{bartumeus_helical_2003}%
  \BibitemOpen
  \bibfield  {author} {\bibinfo {author} {\bibfnamefont {F.}~\bibnamefont {Bartumeus}}, \bibinfo {author} {\bibfnamefont {F.}~\bibnamefont {Peters}}, \bibinfo {author} {\bibfnamefont {S.}~\bibnamefont {Pueyo}}, \bibinfo {author} {\bibfnamefont {C.}~\bibnamefont {Marrasé}},\ and\ \bibinfo {author} {\bibfnamefont {J.}~\bibnamefont {Catalan}},\ }\bibfield  {title} {{\selectlanguage {english}\bibinfo {title} {Helical {Lévy} walks: {Adjusting} searching statistics to resource availability in microzooplankton}},\ }\href {https://doi.org/10.1073/pnas.2137243100} {\bibfield  {journal} {\bibinfo  {journal} {Proceedings of the National Academy of Sciences}\ }\textbf {\bibinfo {volume} {100}},\ \bibinfo {pages} {12771} (\bibinfo {year} {2003})}\BibitemShut {NoStop}%
\bibitem [{\citenamefont {Schmitt}\ and\ \citenamefont {Seuront}(2001)}]{schmitt_multifractal_2001}%
  \BibitemOpen
  \bibfield  {author} {\bibinfo {author} {\bibfnamefont {F.~G.}\ \bibnamefont {Schmitt}}\ and\ \bibinfo {author} {\bibfnamefont {L.}~\bibnamefont {Seuront}},\ }\bibfield  {title} {{\selectlanguage {english}\bibinfo {title} {Multifractal random walk in copepod behavior}},\ }\href {https://doi.org/10.1016/S0378-4371(01)00429-0} {\bibfield  {journal} {\bibinfo  {journal} {Physica A: Statistical Mechanics and its Applications}\ }\textbf {\bibinfo {volume} {301}},\ \bibinfo {pages} {375} (\bibinfo {year} {2001})}\BibitemShut {NoStop}%
\bibitem [{\citenamefont {Chenouard}\ \emph {et~al.}(2014)\citenamefont {Chenouard}, \citenamefont {Smal}, \citenamefont {de~Chaumont}, \citenamefont {Maška}, \citenamefont {Sbalzarini}, \citenamefont {Gong}, \citenamefont {Cardinale}, \citenamefont {Carthel}, \citenamefont {Coraluppi}, \citenamefont {Winter}, \citenamefont {Cohen}, \citenamefont {Godinez}, \citenamefont {Rohr}, \citenamefont {Kalaidzidis}, \citenamefont {Liang}, \citenamefont {Duncan}, \citenamefont {Shen}, \citenamefont {Xu}, \citenamefont {Magnusson}, \citenamefont {Jaldén}, \citenamefont {Blau}, \citenamefont {Paul-Gilloteaux}, \citenamefont {Roudot}, \citenamefont {Kervrann}, \citenamefont {Waharte}, \citenamefont {Tinevez}, \citenamefont {Shorte}, \citenamefont {Willemse}, \citenamefont {Celler}, \citenamefont {van Wezel}, \citenamefont {Dan}, \citenamefont {Tsai}, \citenamefont {de~Solórzano}, \citenamefont {Olivo-Marin},\ and\ \citenamefont {Meijering}}]{chenouard_objective_2014}%
  \BibitemOpen
  \bibfield  {author} {\bibinfo {author} {\bibfnamefont {N.}~\bibnamefont {Chenouard}}, \bibinfo {author} {\bibfnamefont {I.}~\bibnamefont {Smal}}, \bibinfo {author} {\bibfnamefont {F.}~\bibnamefont {de~Chaumont}}, \bibinfo {author} {\bibfnamefont {M.}~\bibnamefont {Maška}}, \bibinfo {author} {\bibfnamefont {I.~F.}\ \bibnamefont {Sbalzarini}}, \bibinfo {author} {\bibfnamefont {Y.}~\bibnamefont {Gong}}, \bibinfo {author} {\bibfnamefont {J.}~\bibnamefont {Cardinale}}, \bibinfo {author} {\bibfnamefont {C.}~\bibnamefont {Carthel}}, \bibinfo {author} {\bibfnamefont {S.}~\bibnamefont {Coraluppi}}, \bibinfo {author} {\bibfnamefont {M.}~\bibnamefont {Winter}}, \bibinfo {author} {\bibfnamefont {A.~R.}\ \bibnamefont {Cohen}}, \bibinfo {author} {\bibfnamefont {W.~J.}\ \bibnamefont {Godinez}}, \bibinfo {author} {\bibfnamefont {K.}~\bibnamefont {Rohr}}, \bibinfo {author} {\bibfnamefont {Y.}~\bibnamefont {Kalaidzidis}}, \bibinfo {author} {\bibfnamefont {L.}~\bibnamefont {Liang}}, \bibinfo {author} {\bibfnamefont {J.}~\bibnamefont {Duncan}}, \bibinfo {author} {\bibfnamefont {H.}~\bibnamefont {Shen}}, \bibinfo {author} {\bibfnamefont {Y.}~\bibnamefont {Xu}}, \bibinfo {author} {\bibfnamefont {K.~E.~G.}\ \bibnamefont {Magnusson}}, \bibinfo {author} {\bibfnamefont {J.}~\bibnamefont {Jaldén}}, \bibinfo {author} {\bibfnamefont {H.~M.}\ \bibnamefont {Blau}}, \bibinfo {author} {\bibfnamefont {P.}~\bibnamefont {Paul-Gilloteaux}}, \bibinfo {author} {\bibfnamefont {P.}~\bibnamefont {Roudot}}, \bibinfo {author} {\bibfnamefont {C.}~\bibnamefont {Kervrann}}, \bibinfo {author} {\bibfnamefont {F.}~\bibnamefont {Waharte}}, \bibinfo {author} {\bibfnamefont {J.-Y.}\ \bibnamefont {Tinevez}}, \bibinfo {author} {\bibfnamefont {S.~L.}\ \bibnamefont {Shorte}}, \bibinfo {author} {\bibfnamefont {J.}~\bibnamefont {Willemse}}, \bibinfo {author} {\bibfnamefont {K.}~\bibnamefont {Celler}}, \bibinfo {author} {\bibfnamefont {G.~P.}\ \bibnamefont {van Wezel}}, \bibinfo {author} {\bibfnamefont {H.-W.}\ \bibnamefont {Dan}}, \bibinfo {author} {\bibfnamefont {Y.-S.}\ \bibnamefont {Tsai}}, \bibinfo {author} {\bibfnamefont {C.~O.}\ \bibnamefont {de~Solórzano}}, \bibinfo {author} {\bibfnamefont {J.-C.}\ \bibnamefont {Olivo-Marin}},\ and\ \bibinfo {author} {\bibfnamefont {E.}~\bibnamefont {Meijering}},\ }\bibfield  {title} {{\selectlanguage {english}\bibinfo {title} {Objective comparison of particle tracking methods}},\ }\href {https://doi.org/10.1038/nmeth.2808} {\bibfield  {journal} {\bibinfo  {journal} {Nature Methods}\ }\textbf {\bibinfo {volume} {11}},\ \bibinfo {pages} {281} (\bibinfo {year} {2014})},\ \bibinfo {note} {number: 3 Publisher: Nature Publishing Group}\BibitemShut {NoStop}%
\bibitem [{\citenamefont {Ulman}\ \emph {et~al.}(2017)\citenamefont {Ulman}, \citenamefont {Ma{\v{s}}ka}, \citenamefont {Magnusson}, \citenamefont {Ronneberger}, \citenamefont {Haubold}, \citenamefont {Harder}, \citenamefont {Matula}, \citenamefont {Matula}, \citenamefont {Svoboda}, \citenamefont {Radojevic} \emph {et~al.}}]{ulman_objective_2017}%
  \BibitemOpen
  \bibfield  {author} {\bibinfo {author} {\bibfnamefont {V.}~\bibnamefont {Ulman}}, \bibinfo {author} {\bibfnamefont {M.}~\bibnamefont {Ma{\v{s}}ka}}, \bibinfo {author} {\bibfnamefont {K.~E.}\ \bibnamefont {Magnusson}}, \bibinfo {author} {\bibfnamefont {O.}~\bibnamefont {Ronneberger}}, \bibinfo {author} {\bibfnamefont {C.}~\bibnamefont {Haubold}}, \bibinfo {author} {\bibfnamefont {N.}~\bibnamefont {Harder}}, \bibinfo {author} {\bibfnamefont {P.}~\bibnamefont {Matula}}, \bibinfo {author} {\bibfnamefont {P.}~\bibnamefont {Matula}}, \bibinfo {author} {\bibfnamefont {D.}~\bibnamefont {Svoboda}}, \bibinfo {author} {\bibfnamefont {M.}~\bibnamefont {Radojevic}}, \emph {et~al.},\ }\bibfield  {title} {\bibinfo {title} {An objective comparison of cell-tracking algorithms},\ }\href@noop {} {\bibfield  {journal} {\bibinfo  {journal} {Nature methods}\ }\textbf {\bibinfo {volume} {14}},\ \bibinfo {pages} {1141} (\bibinfo {year} {2017})}\BibitemShut {NoStop}%
\bibitem [{\citenamefont {Godinez}\ and\ \citenamefont {Rohr}(2015)}]{godinez_tracking_2015}%
  \BibitemOpen
  \bibfield  {author} {\bibinfo {author} {\bibfnamefont {W.~J.}\ \bibnamefont {Godinez}}\ and\ \bibinfo {author} {\bibfnamefont {K.}~\bibnamefont {Rohr}},\ }\bibfield  {title} {{\selectlanguage {english}\bibinfo {title} {Tracking multiple particles in fluorescence time-lapse microscopy images via probabilistic data association}},\ }\href {https://doi.org/10.1109/TMI.2014.2359541} {\bibfield  {journal} {\bibinfo  {journal} {IEEE transactions on medical imaging}\ }\textbf {\bibinfo {volume} {34}},\ \bibinfo {pages} {415} (\bibinfo {year} {2015})}\BibitemShut {NoStop}%
\bibitem [{\citenamefont {Spilger}\ \emph {et~al.}(2020)\citenamefont {Spilger}, \citenamefont {Imle}, \citenamefont {Lee}, \citenamefont {Müller}, \citenamefont {Fackler}, \citenamefont {Bartenschlager},\ and\ \citenamefont {Rohr}}]{spilger_recurrent_2020}%
  \BibitemOpen
  \bibfield  {author} {\bibinfo {author} {\bibfnamefont {R.}~\bibnamefont {Spilger}}, \bibinfo {author} {\bibfnamefont {A.}~\bibnamefont {Imle}}, \bibinfo {author} {\bibfnamefont {J.-Y.}\ \bibnamefont {Lee}}, \bibinfo {author} {\bibfnamefont {B.}~\bibnamefont {Müller}}, \bibinfo {author} {\bibfnamefont {O.~T.}\ \bibnamefont {Fackler}}, \bibinfo {author} {\bibfnamefont {R.}~\bibnamefont {Bartenschlager}},\ and\ \bibinfo {author} {\bibfnamefont {K.}~\bibnamefont {Rohr}},\ }\bibfield  {title} {\bibinfo {title} {A {Recurrent} {Neural} {Network} for {Particle} {Tracking} in {Microscopy} {Images} {Using} {Future} {Information}, {Track} {Hypotheses}, and {Multiple} {Detections}},\ }\href {https://doi.org/10.1109/TIP.2020.2964515} {\bibfield  {journal} {\bibinfo  {journal} {IEEE Transactions on Image Processing}\ }\textbf {\bibinfo {volume} {29}},\ \bibinfo {pages} {3681} (\bibinfo {year} {2020})}\BibitemShut {NoStop}%
\bibitem [{\citenamefont {Spilger}\ \emph {et~al.}(2021)\citenamefont {Spilger}, \citenamefont {Lee}, \citenamefont {Chagin}, \citenamefont {Schermelleh}, \citenamefont {Cardoso}, \citenamefont {Bartenschlager},\ and\ \citenamefont {Rohr}}]{spilger_deep_2021}%
  \BibitemOpen
  \bibfield  {author} {\bibinfo {author} {\bibfnamefont {R.}~\bibnamefont {Spilger}}, \bibinfo {author} {\bibfnamefont {J.-Y.}\ \bibnamefont {Lee}}, \bibinfo {author} {\bibfnamefont {V.~O.}\ \bibnamefont {Chagin}}, \bibinfo {author} {\bibfnamefont {L.}~\bibnamefont {Schermelleh}}, \bibinfo {author} {\bibfnamefont {M.~C.}\ \bibnamefont {Cardoso}}, \bibinfo {author} {\bibfnamefont {R.}~\bibnamefont {Bartenschlager}},\ and\ \bibinfo {author} {\bibfnamefont {K.}~\bibnamefont {Rohr}},\ }\bibfield  {title} {{\selectlanguage {english}\bibinfo {title} {Deep probabilistic tracking of particles in fluorescence microscopy images}},\ }\href {https://doi.org/10.1016/j.media.2021.102128} {\bibfield  {journal} {\bibinfo  {journal} {Medical Image Analysis}\ }\textbf {\bibinfo {volume} {72}},\ \bibinfo {pages} {102128} (\bibinfo {year} {2021})}\BibitemShut {NoStop}%
\bibitem [{\citenamefont {Yao}\ \emph {et~al.}(2020)\citenamefont {Yao}, \citenamefont {Smal}, \citenamefont {Grigoriev}, \citenamefont {Akhmanova},\ and\ \citenamefont {Meijering}}]{yao_deep-learning_2020}%
  \BibitemOpen
  \bibfield  {author} {\bibinfo {author} {\bibfnamefont {Y.}~\bibnamefont {Yao}}, \bibinfo {author} {\bibfnamefont {I.}~\bibnamefont {Smal}}, \bibinfo {author} {\bibfnamefont {I.}~\bibnamefont {Grigoriev}}, \bibinfo {author} {\bibfnamefont {A.}~\bibnamefont {Akhmanova}},\ and\ \bibinfo {author} {\bibfnamefont {E.}~\bibnamefont {Meijering}},\ }\bibfield  {title} {{\selectlanguage {english}\bibinfo {title} {Deep-learning method for data association in particle tracking}},\ }\href {https://doi.org/10.1093/bioinformatics/btaa597} {\bibfield  {journal} {\bibinfo  {journal} {Bioinformatics (Oxford, England)}\ }\textbf {\bibinfo {volume} {36}},\ \bibinfo {pages} {4935} (\bibinfo {year} {2020})}\BibitemShut {NoStop}%
\bibitem [{\citenamefont {Pineda}\ \emph {et~al.}(2023)\citenamefont {Pineda}, \citenamefont {Midtvedt}, \citenamefont {Bachimanchi}, \citenamefont {Noé}, \citenamefont {Midtvedt}, \citenamefont {Volpe},\ and\ \citenamefont {Manzo}}]{pineda_geometric_2023}%
  \BibitemOpen
  \bibfield  {author} {\bibinfo {author} {\bibfnamefont {J.}~\bibnamefont {Pineda}}, \bibinfo {author} {\bibfnamefont {B.}~\bibnamefont {Midtvedt}}, \bibinfo {author} {\bibfnamefont {H.}~\bibnamefont {Bachimanchi}}, \bibinfo {author} {\bibfnamefont {S.}~\bibnamefont {Noé}}, \bibinfo {author} {\bibfnamefont {D.}~\bibnamefont {Midtvedt}}, \bibinfo {author} {\bibfnamefont {G.}~\bibnamefont {Volpe}},\ and\ \bibinfo {author} {\bibfnamefont {C.}~\bibnamefont {Manzo}},\ }\bibfield  {title} {{\selectlanguage {english}\bibinfo {title} {Geometric deep learning reveals the spatiotemporal features of microscopic motion}},\ }\href {https://doi.org/10.1038/s42256-022-00595-0} {\bibfield  {journal} {\bibinfo  {journal} {Nature Machine Intelligence}\ }\textbf {\bibinfo {volume} {5}},\ \bibinfo {pages} {71} (\bibinfo {year} {2023})},\ \bibinfo {note} {number: 1 Publisher: Nature Publishing Group}\BibitemShut {NoStop}%
\bibitem [{\citenamefont {Ying}\ \emph {et~al.}(2021)\citenamefont {Ying}, \citenamefont {Cai}, \citenamefont {Luo}, \citenamefont {Zheng}, \citenamefont {Ke}, \citenamefont {He}, \citenamefont {Shen},\ and\ \citenamefont {Liu}}]{ying_transformers_2021}%
  \BibitemOpen
  \bibfield  {author} {\bibinfo {author} {\bibfnamefont {C.}~\bibnamefont {Ying}}, \bibinfo {author} {\bibfnamefont {T.}~\bibnamefont {Cai}}, \bibinfo {author} {\bibfnamefont {S.}~\bibnamefont {Luo}}, \bibinfo {author} {\bibfnamefont {S.}~\bibnamefont {Zheng}}, \bibinfo {author} {\bibfnamefont {G.}~\bibnamefont {Ke}}, \bibinfo {author} {\bibfnamefont {D.}~\bibnamefont {He}}, \bibinfo {author} {\bibfnamefont {Y.}~\bibnamefont {Shen}},\ and\ \bibinfo {author} {\bibfnamefont {T.-Y.}\ \bibnamefont {Liu}},\ }\href {https://doi.org/10.48550/arXiv.2106.05234} {\bibinfo {title} {Do {Transformers} {Really} {Perform} {Bad} for {Graph} {Representation}?}} (\bibinfo {year} {2021}),\ \bibinfo {note} {arXiv:2106.05234 [cs]}\BibitemShut {NoStop}%
\bibitem [{\citenamefont {Vaswani}\ \emph {et~al.}(2017)\citenamefont {Vaswani}, \citenamefont {Shazeer}, \citenamefont {Parmar}, \citenamefont {Uszkoreit}, \citenamefont {Jones}, \citenamefont {Gomez}, \citenamefont {Kaiser},\ and\ \citenamefont {Polosukhin}}]{vaswani2017attention}%
  \BibitemOpen
  \bibfield  {author} {\bibinfo {author} {\bibfnamefont {A.}~\bibnamefont {Vaswani}}, \bibinfo {author} {\bibfnamefont {N.}~\bibnamefont {Shazeer}}, \bibinfo {author} {\bibfnamefont {N.}~\bibnamefont {Parmar}}, \bibinfo {author} {\bibfnamefont {J.}~\bibnamefont {Uszkoreit}}, \bibinfo {author} {\bibfnamefont {L.}~\bibnamefont {Jones}}, \bibinfo {author} {\bibfnamefont {A.~N.}\ \bibnamefont {Gomez}}, \bibinfo {author} {\bibfnamefont {{\L}.}~\bibnamefont {Kaiser}},\ and\ \bibinfo {author} {\bibfnamefont {I.}~\bibnamefont {Polosukhin}},\ }\bibfield  {title} {\bibinfo {title} {Attention is all you need},\ }\href@noop {} {\bibfield  {journal} {\bibinfo  {journal} {Advances in neural information processing systems}\ }\textbf {\bibinfo {volume} {30}} (\bibinfo {year} {2017})}\BibitemShut {NoStop}%
\bibitem [{\citenamefont {Dosovitskiy}\ \emph {et~al.}(2020)\citenamefont {Dosovitskiy}, \citenamefont {Beyer}, \citenamefont {Kolesnikov}, \citenamefont {Weissenborn}, \citenamefont {Zhai}, \citenamefont {Unterthiner}, \citenamefont {Dehghani}, \citenamefont {Minderer}, \citenamefont {Heigold}, \citenamefont {Gelly} \emph {et~al.}}]{dosovitskiy2020image}%
  \BibitemOpen
  \bibfield  {author} {\bibinfo {author} {\bibfnamefont {A.}~\bibnamefont {Dosovitskiy}}, \bibinfo {author} {\bibfnamefont {L.}~\bibnamefont {Beyer}}, \bibinfo {author} {\bibfnamefont {A.}~\bibnamefont {Kolesnikov}}, \bibinfo {author} {\bibfnamefont {D.}~\bibnamefont {Weissenborn}}, \bibinfo {author} {\bibfnamefont {X.}~\bibnamefont {Zhai}}, \bibinfo {author} {\bibfnamefont {T.}~\bibnamefont {Unterthiner}}, \bibinfo {author} {\bibfnamefont {M.}~\bibnamefont {Dehghani}}, \bibinfo {author} {\bibfnamefont {M.}~\bibnamefont {Minderer}}, \bibinfo {author} {\bibfnamefont {G.}~\bibnamefont {Heigold}}, \bibinfo {author} {\bibfnamefont {S.}~\bibnamefont {Gelly}}, \emph {et~al.},\ }\bibfield  {title} {\bibinfo {title} {An image is worth 16x16 words: Transformers for image recognition at scale},\ }\href@noop {} {\bibfield  {journal} {\bibinfo  {journal} {arXiv preprint arXiv:2010.11929}\ } (\bibinfo {year} {2020})}\BibitemShut {NoStop}%
\bibitem [{\citenamefont {Caron}\ \emph {et~al.}(2021)\citenamefont {Caron}, \citenamefont {Touvron}, \citenamefont {Misra}, \citenamefont {J{\'e}gou}, \citenamefont {Mairal}, \citenamefont {Bojanowski},\ and\ \citenamefont {Joulin}}]{caron2021emerging}%
  \BibitemOpen
  \bibfield  {author} {\bibinfo {author} {\bibfnamefont {M.}~\bibnamefont {Caron}}, \bibinfo {author} {\bibfnamefont {H.}~\bibnamefont {Touvron}}, \bibinfo {author} {\bibfnamefont {I.}~\bibnamefont {Misra}}, \bibinfo {author} {\bibfnamefont {H.}~\bibnamefont {J{\'e}gou}}, \bibinfo {author} {\bibfnamefont {J.}~\bibnamefont {Mairal}}, \bibinfo {author} {\bibfnamefont {P.}~\bibnamefont {Bojanowski}},\ and\ \bibinfo {author} {\bibfnamefont {A.}~\bibnamefont {Joulin}},\ }\bibfield  {title} {\bibinfo {title} {Emerging properties in self-supervised vision transformers},\ }in\ \href@noop {} {\emph {\bibinfo {booktitle} {Proceedings of the IEEE/CVF international conference on computer vision}}}\ (\bibinfo {year} {2021})\ pp.\ \bibinfo {pages} {9650--9660}\BibitemShut {NoStop}%
\bibitem [{\citenamefont {Lee}\ \emph {et~al.}(2021)\citenamefont {Lee}, \citenamefont {Jeong},\ and\ \citenamefont {Ko}}]{lee_graph_2021}%
  \BibitemOpen
  \bibfield  {author} {\bibinfo {author} {\bibfnamefont {J.}~\bibnamefont {Lee}}, \bibinfo {author} {\bibfnamefont {M.}~\bibnamefont {Jeong}},\ and\ \bibinfo {author} {\bibfnamefont {B.~C.}\ \bibnamefont {Ko}},\ }\bibfield  {title} {\bibinfo {title} {Graph {Convolution} {Neural} {Network}-{Based} {Data} {Association} for {Online} {Multi}-{Object} {Tracking}},\ }\href {https://doi.org/10.1109/ACCESS.2021.3105118} {\bibfield  {journal} {\bibinfo  {journal} {IEEE Access}\ }\textbf {\bibinfo {volume} {9}},\ \bibinfo {pages} {114535} (\bibinfo {year} {2021})},\ \bibinfo {note} {conference Name: IEEE Access}\BibitemShut {NoStop}%
\bibitem [{\citenamefont {Kipf}\ and\ \citenamefont {Welling}(2017)}]{kipf_semi-supervised_2017}%
  \BibitemOpen
  \bibfield  {author} {\bibinfo {author} {\bibfnamefont {T.~N.}\ \bibnamefont {Kipf}}\ and\ \bibinfo {author} {\bibfnamefont {M.}~\bibnamefont {Welling}},\ }\href {https://doi.org/10.48550/arXiv.1609.02907} {\bibinfo {title} {Semi-{Supervised} {Classification} with {Graph} {Convolutional} {Networks}}} (\bibinfo {year} {2017}),\ \bibinfo {note} {arXiv:1609.02907 [cs, stat]}\BibitemShut {NoStop}%
\bibitem [{\citenamefont {Qiu}\ \emph {et~al.}(2022{\natexlab{a}})\citenamefont {Qiu}, \citenamefont {Mousavi}, \citenamefont {Zhao},\ and\ \citenamefont {Gustavsson}}]{qiu_active_2022}%
  \BibitemOpen
  \bibfield  {author} {\bibinfo {author} {\bibfnamefont {J.}~\bibnamefont {Qiu}}, \bibinfo {author} {\bibfnamefont {N.}~\bibnamefont {Mousavi}}, \bibinfo {author} {\bibfnamefont {L.}~\bibnamefont {Zhao}},\ and\ \bibinfo {author} {\bibfnamefont {K.}~\bibnamefont {Gustavsson}},\ }\bibfield  {title} {\bibinfo {title} {Active gyrotactic stability of microswimmers using hydromechanical signals},\ }\bibfield  {journal} {\bibinfo  {journal} {Physical Review Fluids}\ }\textbf {\bibinfo {volume} {7}},\ \href {https://doi.org/10.1103/PhysRevFluids.7.014311} {10.1103/PhysRevFluids.7.014311} (\bibinfo {year} {2022}{\natexlab{a}})\BibitemShut {NoStop}%
\bibitem [{\citenamefont {Gustavsson}\ \emph {et~al.}(2017)\citenamefont {Gustavsson}, \citenamefont {Biferale}, \citenamefont {Celani},\ and\ \citenamefont {Colabrese}}]{gustavsson_finding_2017}%
  \BibitemOpen
  \bibfield  {author} {\bibinfo {author} {\bibfnamefont {K.}~\bibnamefont {Gustavsson}}, \bibinfo {author} {\bibfnamefont {L.}~\bibnamefont {Biferale}}, \bibinfo {author} {\bibfnamefont {A.}~\bibnamefont {Celani}},\ and\ \bibinfo {author} {\bibfnamefont {S.}~\bibnamefont {Colabrese}},\ }\bibfield  {title} {{\selectlanguage {english}\bibinfo {title} {Finding efficient swimming strategies in a three-dimensional chaotic flow by reinforcement learning}},\ }\href {https://doi.org/10.1140/epje/i2017-11602-9} {\bibfield  {journal} {\bibinfo  {journal} {The European Physical Journal E}\ }\textbf {\bibinfo {volume} {40}},\ \bibinfo {pages} {110} (\bibinfo {year} {2017})}\BibitemShut {NoStop}%
\bibitem [{\citenamefont {Colabrese}\ \emph {et~al.}(2017)\citenamefont {Colabrese}, \citenamefont {Gustavsson}, \citenamefont {Celani},\ and\ \citenamefont {Biferale}}]{colabrese_flow_2017}%
  \BibitemOpen
  \bibfield  {author} {\bibinfo {author} {\bibfnamefont {S.}~\bibnamefont {Colabrese}}, \bibinfo {author} {\bibfnamefont {K.}~\bibnamefont {Gustavsson}}, \bibinfo {author} {\bibfnamefont {A.}~\bibnamefont {Celani}},\ and\ \bibinfo {author} {\bibfnamefont {L.}~\bibnamefont {Biferale}},\ }\bibfield  {title} {\bibinfo {title} {Flow {Navigation} by {Smart} {Microswimmers} via {Reinforcement} {Learning}},\ }\href {https://doi.org/10.1103/PhysRevLett.118.158004} {\bibfield  {journal} {\bibinfo  {journal} {Physical Review Letters}\ }\textbf {\bibinfo {volume} {118}},\ \bibinfo {pages} {158004} (\bibinfo {year} {2017})},\ \bibinfo {note} {publisher: American Physical Society}\BibitemShut {NoStop}%
\bibitem [{\citenamefont {Gunnarson}\ \emph {et~al.}(2021)\citenamefont {Gunnarson}, \citenamefont {Mandralis}, \citenamefont {Novati}, \citenamefont {Koumoutsakos},\ and\ \citenamefont {Dabiri}}]{gunnarson_learning_2021}%
  \BibitemOpen
  \bibfield  {author} {\bibinfo {author} {\bibfnamefont {P.}~\bibnamefont {Gunnarson}}, \bibinfo {author} {\bibfnamefont {I.}~\bibnamefont {Mandralis}}, \bibinfo {author} {\bibfnamefont {G.}~\bibnamefont {Novati}}, \bibinfo {author} {\bibfnamefont {P.}~\bibnamefont {Koumoutsakos}},\ and\ \bibinfo {author} {\bibfnamefont {J.~O.}\ \bibnamefont {Dabiri}},\ }\bibfield  {title} {{\selectlanguage {english}\bibinfo {title} {Learning efficient navigation in vortical flow fields}},\ }\href {https://doi.org/10.1038/s41467-021-27015-y} {\bibfield  {journal} {\bibinfo  {journal} {Nature Communications}\ }\textbf {\bibinfo {volume} {12}},\ \bibinfo {pages} {7143} (\bibinfo {year} {2021})},\ \bibinfo {note} {number: 1 Publisher: Nature Publishing Group}\BibitemShut {NoStop}%
\bibitem [{\citenamefont {Alageshan}\ \emph {et~al.}(2020)\citenamefont {Alageshan}, \citenamefont {Verma}, \citenamefont {Bec},\ and\ \citenamefont {Pandit}}]{alageshan_machine_2020}%
  \BibitemOpen
  \bibfield  {author} {\bibinfo {author} {\bibfnamefont {J.~K.}\ \bibnamefont {Alageshan}}, \bibinfo {author} {\bibfnamefont {A.~K.}\ \bibnamefont {Verma}}, \bibinfo {author} {\bibfnamefont {J.}~\bibnamefont {Bec}},\ and\ \bibinfo {author} {\bibfnamefont {R.}~\bibnamefont {Pandit}},\ }\bibfield  {title} {\bibinfo {title} {Machine learning strategies for path-planning microswimmers in turbulent flows},\ }\href {https://doi.org/10.1103/PhysRevE.101.043110} {\bibfield  {journal} {\bibinfo  {journal} {Physical Review E}\ }\textbf {\bibinfo {volume} {101}},\ \bibinfo {pages} {043110} (\bibinfo {year} {2020})},\ \bibinfo {note} {publisher: American Physical Society}\BibitemShut {NoStop}%
\bibitem [{\citenamefont {Qiu}\ \emph {et~al.}(2022{\natexlab{b}})\citenamefont {Qiu}, \citenamefont {Mousavi}, \citenamefont {Gustavsson}, \citenamefont {Xu}, \citenamefont {Mehlig},\ and\ \citenamefont {Zhao}}]{qiu_navigation_2022}%
  \BibitemOpen
  \bibfield  {author} {\bibinfo {author} {\bibfnamefont {J.}~\bibnamefont {Qiu}}, \bibinfo {author} {\bibfnamefont {N.}~\bibnamefont {Mousavi}}, \bibinfo {author} {\bibfnamefont {K.}~\bibnamefont {Gustavsson}}, \bibinfo {author} {\bibfnamefont {C.}~\bibnamefont {Xu}}, \bibinfo {author} {\bibfnamefont {B.}~\bibnamefont {Mehlig}},\ and\ \bibinfo {author} {\bibfnamefont {L.}~\bibnamefont {Zhao}},\ }\bibfield  {title} {{\selectlanguage {english}\bibinfo {title} {Navigation of micro-swimmers in steady flow: the importance of symmetries}},\ }\href {https://doi.org/10.1017/jfm.2021.978} {\bibfield  {journal} {\bibinfo  {journal} {Journal of Fluid Mechanics}\ }\textbf {\bibinfo {volume} {932}},\ \bibinfo {pages} {A10} (\bibinfo {year} {2022}{\natexlab{b}})},\ \bibinfo {note} {publisher: Cambridge University Press}\BibitemShut {NoStop}%
\bibitem [{\citenamefont {D{\'\i}az}\ \emph {et~al.}(2019)\citenamefont {D{\'\i}az}, \citenamefont {{\'A}lvarez}, \citenamefont {Varela}, \citenamefont {P{\'e}rez-Santos}, \citenamefont {D{\'\i}az}, \citenamefont {Molinet}, \citenamefont {Seguel}, \citenamefont {Aguilera-Belmonte}, \citenamefont {Guzm{\'a}n}, \citenamefont {Uribe} \emph {et~al.}}]{diaz2019impacts}%
  \BibitemOpen
  \bibfield  {author} {\bibinfo {author} {\bibfnamefont {P.~A.}\ \bibnamefont {D{\'\i}az}}, \bibinfo {author} {\bibfnamefont {G.}~\bibnamefont {{\'A}lvarez}}, \bibinfo {author} {\bibfnamefont {D.}~\bibnamefont {Varela}}, \bibinfo {author} {\bibfnamefont {I.}~\bibnamefont {P{\'e}rez-Santos}}, \bibinfo {author} {\bibfnamefont {M.}~\bibnamefont {D{\'\i}az}}, \bibinfo {author} {\bibfnamefont {C.}~\bibnamefont {Molinet}}, \bibinfo {author} {\bibfnamefont {M.}~\bibnamefont {Seguel}}, \bibinfo {author} {\bibfnamefont {A.}~\bibnamefont {Aguilera-Belmonte}}, \bibinfo {author} {\bibfnamefont {L.}~\bibnamefont {Guzm{\'a}n}}, \bibinfo {author} {\bibfnamefont {E.}~\bibnamefont {Uribe}}, \emph {et~al.},\ }\bibfield  {title} {\bibinfo {title} {Impacts of harmful algal blooms on the aquaculture industry: Chile as a case study},\ }\href {https://doi.org/10.1127/pip/2019/0081} {\bibfield  {journal} {\bibinfo  {journal} {Perspect. Phycol}\ }\textbf {\bibinfo {volume} {6}},\ \bibinfo {pages} {39} (\bibinfo {year} {2019})}\BibitemShut {NoStop}%
\bibitem [{\citenamefont {Lenzen}\ \emph {et~al.}(2021)\citenamefont {Lenzen}, \citenamefont {Li},\ and\ \citenamefont {Murray}}]{lenzen2021impacts}%
  \BibitemOpen
  \bibfield  {author} {\bibinfo {author} {\bibfnamefont {M.}~\bibnamefont {Lenzen}}, \bibinfo {author} {\bibfnamefont {M.}~\bibnamefont {Li}},\ and\ \bibinfo {author} {\bibfnamefont {S.~A.}\ \bibnamefont {Murray}},\ }\bibfield  {title} {\bibinfo {title} {Impacts of harmful algal blooms on marine aquaculture in a low-carbon future},\ }\href {https://doi.org/10.1016/j.hal.2021.102143} {\bibfield  {journal} {\bibinfo  {journal} {Harmful Algae}\ }\textbf {\bibinfo {volume} {110}},\ \bibinfo {pages} {102143} (\bibinfo {year} {2021})}\BibitemShut {NoStop}%
\bibitem [{\citenamefont {Franks}(2002)}]{franks2002npz}%
  \BibitemOpen
  \bibfield  {author} {\bibinfo {author} {\bibfnamefont {P.~J.}\ \bibnamefont {Franks}},\ }\bibfield  {title} {\bibinfo {title} {Npz models of plankton dynamics: their construction, coupling to physics, and application},\ }\href {https://doi.org/10.1023/A:1015874028196} {\bibfield  {journal} {\bibinfo  {journal} {Journal of Oceanography}\ }\textbf {\bibinfo {volume} {58}},\ \bibinfo {pages} {379} (\bibinfo {year} {2002})}\BibitemShut {NoStop}%
\bibitem [{\citenamefont {Frost}(1987)}]{Frost_1987}%
  \BibitemOpen
  \bibfield  {author} {\bibinfo {author} {\bibfnamefont {B.}~\bibnamefont {Frost}},\ }\bibfield  {title} {\bibinfo {title} {Grazing control of phytoplankton stock in the open subarctic pacific ocean: a model assessing the role of mesozooplankton, particularly the large calanoid copepods neocalanus spp.},\ }\href@noop {} {\bibfield  {journal} {\bibinfo  {journal} {Marine ecology progress series. Oldendorf}\ }\textbf {\bibinfo {volume} {39}},\ \bibinfo {pages} {49} (\bibinfo {year} {1987})}\BibitemShut {NoStop}%
\bibitem [{\citenamefont {Aoki}\ \emph {et~al.}(1999)\citenamefont {Aoki}, \citenamefont {Komatsu},\ and\ \citenamefont {Hwang}}]{Aoki_Komatsu_Hwang_1999}%
  \BibitemOpen
  \bibfield  {author} {\bibinfo {author} {\bibfnamefont {I.}~\bibnamefont {Aoki}}, \bibinfo {author} {\bibfnamefont {T.}~\bibnamefont {Komatsu}},\ and\ \bibinfo {author} {\bibfnamefont {K.}~\bibnamefont {Hwang}},\ }\bibfield  {title} {{\selectlanguage {english}\bibinfo {title} {Prediction of response of zooplankton biomass to climatic and oceanic changes}},\ }\href {https://doi.org/10.1016/S0304-3800(99)00107-6} {\bibfield  {journal} {\bibinfo  {journal} {Ecological Modelling}\ }\textbf {\bibinfo {volume} {120}},\ \bibinfo {pages} {261–270} (\bibinfo {year} {1999})}\BibitemShut {NoStop}%
\bibitem [{\citenamefont {Woodd-Walker}\ \emph {et~al.}(2001)\citenamefont {Woodd-Walker}, \citenamefont {Kingston},\ and\ \citenamefont {Gallienne}}]{Woodd-Walker_Kingston_Gallienne}%
  \BibitemOpen
  \bibfield  {author} {\bibinfo {author} {\bibfnamefont {R.~S.}\ \bibnamefont {Woodd-Walker}}, \bibinfo {author} {\bibfnamefont {K.~S.}\ \bibnamefont {Kingston}},\ and\ \bibinfo {author} {\bibfnamefont {C.~P.}\ \bibnamefont {Gallienne}},\ }\bibfield  {title} {{\selectlanguage {english}\bibinfo {title} {Using neural networks to predict surface zooplankton biomass along a 50°n to 50°s transect of the atlantic}},\ }\href@noop {} {\bibfield  {journal} {\bibinfo  {journal} {Journal of plankton research}\ }\textbf {\bibinfo {volume} {23}},\ \bibinfo {pages} {875} (\bibinfo {year} {2001})}\BibitemShut {NoStop}%
\bibitem [{\citenamefont {Pan}\ \emph {et~al.}(2020)\citenamefont {Pan}, \citenamefont {Vernet}, \citenamefont {Manck}, \citenamefont {Forsch}, \citenamefont {Ekern}, \citenamefont {Mascioni}, \citenamefont {Barbeau}, \citenamefont {Almandoz},\ and\ \citenamefont {Orona}}]{pan2020environmental}%
  \BibitemOpen
  \bibfield  {author} {\bibinfo {author} {\bibfnamefont {B.~J.}\ \bibnamefont {Pan}}, \bibinfo {author} {\bibfnamefont {M.}~\bibnamefont {Vernet}}, \bibinfo {author} {\bibfnamefont {L.}~\bibnamefont {Manck}}, \bibinfo {author} {\bibfnamefont {K.}~\bibnamefont {Forsch}}, \bibinfo {author} {\bibfnamefont {L.}~\bibnamefont {Ekern}}, \bibinfo {author} {\bibfnamefont {M.}~\bibnamefont {Mascioni}}, \bibinfo {author} {\bibfnamefont {K.~A.}\ \bibnamefont {Barbeau}}, \bibinfo {author} {\bibfnamefont {G.~O.}\ \bibnamefont {Almandoz}},\ and\ \bibinfo {author} {\bibfnamefont {A.~J.}\ \bibnamefont {Orona}},\ }\bibfield  {title} {\bibinfo {title} {Environmental drivers of phytoplankton taxonomic composition in an antarctic fjord},\ }\href {https://doi.org/10.1016/j.pocean.2020.102295} {\bibfield  {journal} {\bibinfo  {journal} {Progress in Oceanography}\ }\textbf {\bibinfo {volume} {183}},\ \bibinfo {pages} {102295} (\bibinfo {year} {2020})}\BibitemShut {NoStop}%
\bibitem [{\citenamefont {Chase}\ \emph {et~al.}(2022)\citenamefont {Chase}, \citenamefont {Boss}, \citenamefont {Ha{\"e}ntjens}, \citenamefont {Culhane}, \citenamefont {Roesler},\ and\ \citenamefont {Karp-Boss}}]{chase2022plankton}%
  \BibitemOpen
  \bibfield  {author} {\bibinfo {author} {\bibfnamefont {A.}~\bibnamefont {Chase}}, \bibinfo {author} {\bibfnamefont {E.}~\bibnamefont {Boss}}, \bibinfo {author} {\bibfnamefont {N.}~\bibnamefont {Ha{\"e}ntjens}}, \bibinfo {author} {\bibfnamefont {E.}~\bibnamefont {Culhane}}, \bibinfo {author} {\bibfnamefont {C.}~\bibnamefont {Roesler}},\ and\ \bibinfo {author} {\bibfnamefont {L.}~\bibnamefont {Karp-Boss}},\ }\bibfield  {title} {\bibinfo {title} {Plankton imagery data inform satellite-based estimates of diatom carbon},\ }\href {https://doi.org/10.1029/2022GL098076} {\bibfield  {journal} {\bibinfo  {journal} {Geophysical Research Letters}\ }\textbf {\bibinfo {volume} {49}},\ \bibinfo {pages} {e2022GL098076} (\bibinfo {year} {2022})}\BibitemShut {NoStop}%
\bibitem [{\citenamefont {Pyo}\ \emph {et~al.}(2022)\citenamefont {Pyo}, \citenamefont {Hong}, \citenamefont {Jang}, \citenamefont {Park}, \citenamefont {Park}, \citenamefont {Noh},\ and\ \citenamefont {Cho}}]{Pyo_Hong_Jang_Park_Park_Noh_Cho_2022}%
  \BibitemOpen
  \bibfield  {author} {\bibinfo {author} {\bibfnamefont {J.}~\bibnamefont {Pyo}}, \bibinfo {author} {\bibfnamefont {S.~M.}\ \bibnamefont {Hong}}, \bibinfo {author} {\bibfnamefont {J.}~\bibnamefont {Jang}}, \bibinfo {author} {\bibfnamefont {S.}~\bibnamefont {Park}}, \bibinfo {author} {\bibfnamefont {J.}~\bibnamefont {Park}}, \bibinfo {author} {\bibfnamefont {J.~H.}\ \bibnamefont {Noh}},\ and\ \bibinfo {author} {\bibfnamefont {K.~H.}\ \bibnamefont {Cho}},\ }\bibfield  {title} {\bibinfo {title} {Drone-borne sensing of major and accessory pigments in algae using deep learning modeling},\ }\href {https://doi.org/10.1080/15481603.2022.2027120} {\bibfield  {journal} {\bibinfo  {journal} {GIScience \& Remote Sensing}\ }\textbf {\bibinfo {volume} {59}},\ \bibinfo {pages} {310–332} (\bibinfo {year} {2022})}\BibitemShut {NoStop}%
\bibitem [{\citenamefont {Sammartino}\ \emph {et~al.}(2018)\citenamefont {Sammartino}, \citenamefont {Marullo}, \citenamefont {Santoleri},\ and\ \citenamefont {Scardi}}]{Sammartino_Marullo_Santoleri_Scardi_2018}%
  \BibitemOpen
  \bibfield  {author} {\bibinfo {author} {\bibfnamefont {M.}~\bibnamefont {Sammartino}}, \bibinfo {author} {\bibfnamefont {S.}~\bibnamefont {Marullo}}, \bibinfo {author} {\bibfnamefont {R.}~\bibnamefont {Santoleri}},\ and\ \bibinfo {author} {\bibfnamefont {M.}~\bibnamefont {Scardi}},\ }\bibfield  {title} {{\selectlanguage {english}\bibinfo {title} {Modelling the vertical distribution of phytoplankton biomass in the mediterranean sea from satellite data: A neural network approach}},\ }\href {https://doi.org/10.3390/rs10101666} {\bibfield  {journal} {\bibinfo  {journal} {Remote Sensing}\ }\textbf {\bibinfo {volume} {10}},\ \bibinfo {pages} {1666} (\bibinfo {year} {2018})}\BibitemShut {NoStop}%
\bibitem [{\citenamefont {Martinez}\ \emph {et~al.}(2020)\citenamefont {Martinez}, \citenamefont {Brini}, \citenamefont {Gorgues}, \citenamefont {Drumetz}, \citenamefont {Roussillon}, \citenamefont {Tandeo}, \citenamefont {Maze},\ and\ \citenamefont {Fablet}}]{Martinez_2020}%
  \BibitemOpen
  \bibfield  {author} {\bibinfo {author} {\bibfnamefont {E.}~\bibnamefont {Martinez}}, \bibinfo {author} {\bibfnamefont {A.}~\bibnamefont {Brini}}, \bibinfo {author} {\bibfnamefont {T.}~\bibnamefont {Gorgues}}, \bibinfo {author} {\bibfnamefont {L.}~\bibnamefont {Drumetz}}, \bibinfo {author} {\bibfnamefont {J.}~\bibnamefont {Roussillon}}, \bibinfo {author} {\bibfnamefont {P.}~\bibnamefont {Tandeo}}, \bibinfo {author} {\bibfnamefont {G.}~\bibnamefont {Maze}},\ and\ \bibinfo {author} {\bibfnamefont {R.}~\bibnamefont {Fablet}},\ }\bibfield  {title} {\bibinfo {title} {Neural network approaches to reconstruct phytoplankton time-series in the global ocean},\ }\bibfield  {journal} {\bibinfo  {journal} {Remote Sensing}\ }\textbf {\bibinfo {volume} {12}},\ \href {https://doi.org/10.3390/rs12244156} {10.3390/rs12244156} (\bibinfo {year} {2020})\BibitemShut {NoStop}%
\bibitem [{\citenamefont {Liu}\ \emph {et~al.}(2022)\citenamefont {Liu}, \citenamefont {He}, \citenamefont {Huang}, \citenamefont {Tang}, \citenamefont {Hu},\ and\ \citenamefont {Xiao}}]{liu2022algal}%
  \BibitemOpen
  \bibfield  {author} {\bibinfo {author} {\bibfnamefont {M.}~\bibnamefont {Liu}}, \bibinfo {author} {\bibfnamefont {J.}~\bibnamefont {He}}, \bibinfo {author} {\bibfnamefont {Y.}~\bibnamefont {Huang}}, \bibinfo {author} {\bibfnamefont {T.}~\bibnamefont {Tang}}, \bibinfo {author} {\bibfnamefont {J.}~\bibnamefont {Hu}},\ and\ \bibinfo {author} {\bibfnamefont {X.}~\bibnamefont {Xiao}},\ }\bibfield  {title} {\bibinfo {title} {Algal bloom forecasting with time-frequency analysis: A hybrid deep learning approach},\ }\href {https://doi.org/10.1029/2022GL098076} {\bibfield  {journal} {\bibinfo  {journal} {Water Research}\ }\textbf {\bibinfo {volume} {219}},\ \bibinfo {pages} {118591} (\bibinfo {year} {2022})}\BibitemShut {NoStop}%
\bibitem [{\citenamefont {Wenxiang}\ \emph {et~al.}(2022)\citenamefont {Wenxiang}, \citenamefont {Caiyun}, \citenamefont {Shaoping},\ and\ \citenamefont {Xueding}}]{wenxiang2022optimization}%
  \BibitemOpen
  \bibfield  {author} {\bibinfo {author} {\bibfnamefont {D.}~\bibnamefont {Wenxiang}}, \bibinfo {author} {\bibfnamefont {Z.}~\bibnamefont {Caiyun}}, \bibinfo {author} {\bibfnamefont {S.}~\bibnamefont {Shaoping}},\ and\ \bibinfo {author} {\bibfnamefont {L.}~\bibnamefont {Xueding}},\ }\bibfield  {title} {\bibinfo {title} {Optimization of deep learning model for coastal chlorophyll a dynamic forecast},\ }\href {https://doi.org/10.1016/j.ecolmodel.2022.109913} {\bibfield  {journal} {\bibinfo  {journal} {Ecological Modelling}\ }\textbf {\bibinfo {volume} {467}},\ \bibinfo {pages} {109913} (\bibinfo {year} {2022})}\BibitemShut {NoStop}%
\bibitem [{\citenamefont {Cichos}\ \emph {et~al.}(2020)\citenamefont {Cichos}, \citenamefont {Gustavsson}, \citenamefont {Mehlig},\ and\ \citenamefont {Volpe}}]{cichos2020machine}%
  \BibitemOpen
  \bibfield  {author} {\bibinfo {author} {\bibfnamefont {F.}~\bibnamefont {Cichos}}, \bibinfo {author} {\bibfnamefont {K.}~\bibnamefont {Gustavsson}}, \bibinfo {author} {\bibfnamefont {B.}~\bibnamefont {Mehlig}},\ and\ \bibinfo {author} {\bibfnamefont {G.}~\bibnamefont {Volpe}},\ }\bibfield  {title} {\bibinfo {title} {Machine learning for active matter},\ }\href@noop {} {\bibfield  {journal} {\bibinfo  {journal} {Nature Machine Intelligence}\ }\textbf {\bibinfo {volume} {2}},\ \bibinfo {pages} {94} (\bibinfo {year} {2020})}\BibitemShut {NoStop}%
\bibitem [{\citenamefont {Baker}(2016)}]{Baker2016}%
  \BibitemOpen
  \bibfield  {author} {\bibinfo {author} {\bibfnamefont {M.}~\bibnamefont {Baker}},\ }\bibfield  {title} {\bibinfo {title} {1,500 scientists lift the lid on reproducibility},\ }\href {https://doi.org/10.1038/533452a} {\bibfield  {journal} {\bibinfo  {journal} {Nature}\ }\textbf {\bibinfo {volume} {533}},\ \bibinfo {pages} {452} (\bibinfo {year} {2016})}\BibitemShut {NoStop}%
\end{thebibliography}%
\vspace{5mm}
\noindent {\bf Competing interests}\\
The authors declare no competing interests. \\

\noindent {\bf Data and code availability}\\
All the relevant source code and the data are made publicly available at the GitHub repository \cite{bachimanchi_planktonreview_2023} \\

\noindent {\bf Author contributions} \\
HB, MIM, CR conceptualised the work, conducted literature research, and wrote the manuscript.
HB developed the code examples and established the GitHub repository.
HB, CR and ES analysed the data, and generated the figures.
PDW, JH, AK, DM, ES and GV provided feedback on the manuscript.
ES and GV provided project oversight, and supervised the project.
\end{document}